\documentclass[referee,useAMS,usenatbib,usegraphicx]{mn2e}
\usepackage{color}

\title[Young Stellar Population of the Bright-Rimmed Clouds BRC 5, BRC 7 and BRC 39]
{Young Stellar Population of Bright-Rimmed Clouds BRC 5, BRC 7 and BRC 39}

\author[Panwar et al.]{Neelam Panwar$^{1,2}$\thanks{E-mail:neelam$\_$$1110$@yahoo.co.in}, W. P. Chen$^{1}$, A. K. Pandey$^3$, M. R. Samal${^4}$, 
K. Ogura${^5}$,
\newauthor D. K. Ojha$^6$, J. Jose$^7$ and B. C. Bhatt$^7$\\
$^1$Graduate Institute of Astronomy, National Central University 300 Jhongda Road, Jhongli 32001, Taiwan\\
$^2$Department of Physics \& Astrophysics, University of Delhi, Delhi - 110007, India\\
$^3$Aryabhatta Research Institute of Observational Sciences (ARIES), Nainital - 263129, India\\
$^4$Laboratoire d'Astrophysique de Marseille -LAM, Universit\'e d'Aix-Marseille \& CNRS, UMR7326 13388 Marseille CEDEX 13 France\\
$^5$Kokugakuin University, Higashi, Shibuya-ku, Tokyo - 1508440, Japan\\
$^6$Tata Institute of Fundamental Research, Mumbai (Bombay) - 400 005, India\\
$^7$Indian Institute of Astrophysics, Koramangala, Bangalore - 560034, India\\}
\begin{document}

\date{}
\pubyear{2014}

\maketitle

\label{firstpage}

\begin{abstract}
Bright-rimmed clouds (BRCs), illuminated and shaped by nearby OB stars, are potential sites of recent/ongoing star formation.  
Here we present an optical and infrared photometric study of three BRCs: BRC\,5, BRC\,7 and BRC\,39 to obtain a census 
of the young stellar population, thereby inferring the star formation scenario, in these regions.  In each BRC, the 
Class~I sources are found to be located mostly near the bright rim or inside the cloud, 
whereas the Class~II sources are preferentially outside, with younger sources 
closer to the rim.  This provides strong support to sequential star formation triggered by radiation driven 
implosion due to the UV radiation. Moreover, each BRC contains a small group of young stars being revealed 
at its head, as the next-generation stars.  In particular, the young stars at the heads of BRC\,5 and 
BRC\,7 are found to be intermediate/high mass stars, which, under proper conditions, may themselves 
trigger further star birth, thereby propagating star formation out to long distances.
\end{abstract}
\begin{keywords}
stars: formation $-$ stars$-$ stars:
pre$-$main$-$sequence $-$ Interstellar medium (ISM): H{\sc ii} regions $-$ ISM: globules
\end{keywords}

\section{Introduction}

The immense stellar winds and UV radiation from massive stars present in a star-forming region have dramatic impact on the immediate vicinity. 
The propagating ionisation fronts can either curb star formation by dispersing surrounding gas, or on the contrary, may 
induce the next-generation star formation. Elmegreen \& Lada (1977) proposed that the expanding ionisation front plays a constructive role 
in inciting a sequence of star formation activities in the neighbourhood. 

Out of a number of processes which induce star formation at the periphery of an H{\sc ii} region, the 
``collect \& collapse'' and ``radiation driven implosion'' (RDI) processes are frequently discussed in the literature. 
In the `collect and collapse' scenario, the material accumulated by the expanding ionisation front and shock front of 
the H{\sc ii} region becomes gravitationally unstable, so it fragments and collapses to form stars (Elmegreen $\&$ Lada 1977). 
In the RDI scenario, a pre-existing dense clump is exposed to the ionising radiation from a massive star or stars, and the 
photoionisation induced shock compresses the head part of the clump to collapse, which consequently leads to 
formation of stars (e.g., Bertoldi 1989, Lefloch $\&$ Lazareff 1995).   
The aligned elongated distribution of young stellar objects (YSOs) in a small molecular cloud 
is considered as an observational signature of the RDI process (Ogura et al. 2002, Lee et al. 2005).
   
Bright-rimmed clouds (BRCs) are small molecular clouds found near the edges of 
evolved H{\sc ii} regions, with the bright rims facing the ionising stars. Their morphology and physical conditions 
match well with the theoretical models of RDI (e.g., Bertoldi 1989), so BRCs are generally believed to be the sites 
of RDI. Many BRCs are indeed associated with Herbig-Haro (HH) objects and 
Infrared Astronomical Satellite (IRAS) point sources indicative of recent star formation.   
Sugitani, Fukui $\&$ Ogura (1991) (hereafter SFO91) and Sugitani $\&$ Ogura (1994) compiled catalogues of 
89 BRCs associated with IRAS point sources in the northern and southern hemispheres. 
Near infrared (NIR) imaging revealed elongated aggregates of YSOs around BRCs with bluer (presumably older) 
stars closer to the ionising source (Sugitani et al. 1995).  It is thus proposed that star formation 
propagates along the axis of the BRC as the ionisation/shock front advances further into the molecular cloud. 
These authors named this phenomenon as the ``small scale sequential star formation'' ({\it S$^4$F}). 

Ogura et al. (2007, hereafter referred as {\it Paper I}) undertook {\it BVI$_{c}$} photometry of four aggregates, BRCs 11NE, 12, 14 and 37, 
and showed quantitatively that the YSOs inside or on the bright rims 
tend to have younger mean ages than those outside, as expected from the $S^4F$ hypothesis. Chauhan et al. (2009, 
hereafter referred as {\it Paper II}) extended the study to a few more BRCs and confirmed the general age gradients. 

Recently, on the basis of  observational characteristics of the BRCs, Morgan et al. (2009) found that not all of 
them are undergoing triggered star formation and refined the SFO91 catalogue by    
classifying BRCs into triggered or non-triggered samples. According to Morgan et al. (2009), the triggered BRCs are expected to follow three observational 
criteria: (1)~presence of the ionised boundary layer (an indicator of strong interaction between the H{\sc ii} region and molecular cloud) 
using 21~cm emission feature from NRAO/VLA Sky Survey (NVSS), (2)~presence of Midcourse Space Experiment (MSX) emission (which traces 
the photon-dominated regions), and (3)~presence of submillimeter emission (which traces the dense cores inside the cloud). 

The present work aims to identify and characterise the young stellar population associated with BRC\,5, BRC\,7 and BRC\,39 
in order (i)~to study the spatial distributions of the YSOs on small scales, (ii)~to examine the age gradients between 
the YSO samples inside/on and outside BRCs, and (iii)~to study the global star formation in the regions. 
Section~2 describes the data sets used in the study.  Section~3 presents a short description of each BRC 
studied in this work. Section~4 describes the methodology to identify and characterise the YSO population using 
near- to far-infrared photometry. Sections~5 describes the determination of various physical parameters of the 
YSOs based on the colour-magnitude diagrams and the fitting of the spectral energy distribution models.  
Section 6 describes star formation scenario in the 
studied BRCs. In section 7 we conclude with the main results of the present work.

\section{Observations and data reductions}
\subsection{Optical observations}
\begin{table*}
\caption{Log of optical observations}
\begin{tabular}{|p{.8in}|p{.8in}|p{2.2in}|p{1.1in}|}
\hline
Region &Telescope  &Filter; exposure time(sec)$\times$No. of frames & Date of observations\\
\hline
BRC 5  &ST, Nainital &V : 300$\times$4;  I$_c$ : 180$\times$4 & 2006.10.17 \\
BRC 7  &ST, Nainital &V : 300$\times$6;   I$_c$ : 300$\times$3 & 2006.10.27 \\
BRC 39 &ST, Nainital &V : 300$\times$2;  I$_c$ : 200$\times$2 & 2006.10.26 \\
BRC 39 &HCT, Hanle   &V : 300$\times$4;  I$_c$ : 200$\times$4 & 2008.10.21 \\
\hline
\end{tabular}
\end{table*}

{\it VI$_{c}$} observations of BRCs~5, 7 and 39 were obtained 
using a 2048 $\times$ 2048 pixel CCD camera mounted at the {f/13} Cassegrain focus of 
the 1.04-m Sampurnanand Telescope (ST) of the Aryabhatta Research Institute of observational 
sciencES (ARIES), Nainital, India. The details of the camera can be found in Pandey et al. (2008), 
and Jose et al. (2012). The CCD with a plate scale of $0^{\prime\prime}.37$/pixel 
covers a field of $\sim$13$\times$13 arcmin$^2$ in the sky.  To improve the signal-to-noise ratio, the 
observations were carried out in a binning mode of $2\times 2$ pixels. During the observations the seeing was $\sim 2^{\prime\prime}$.  The SA98 field (Landolt 1992) 
was observed on 2006 October 26 to standardize the observations of BRC~5 and BRC~39. 
The observations of BRC~7 were standardized by observing the SA92 field on 2009 October 13. 
Deep observations of BRC~39 were carried out using 
the 2-m Himalayan {\it Chandra} Telescope (HCT) on 2008 October 21. The log of the  
observations is tabulated in Table~1. Bias and twilight flat frames were taken during each run.

The pre-processing of the data frames was done using the tasks available under the $IRAF$ data 
reduction software package. The photometric measurements were performed using the $DAOPHOT-II$ 
software package (Stetson 1987). The point spread function was obtained for each frame using several 
uncontaminated stars. The instrumental magnitudes were converted to the standard values by using 
a least-square linear regression procedure (Stetson 1992). The photometric calibration equations 
used  to standardise BRCs 5 and 39 are given in $Paper II$ whereas for BRC~7 are as follows:\\
$v$ = $V$ + (4.660 $\pm$ 0.009) + (0.256 $\pm$ 0.015) $X$ - (0.059 $\pm$ 0.006)$(V-I_c)$,\\
 $i$ = $I_c$ + (5.009 $\pm$ 0.010) + (0.129 $\pm$ 0.016) $X$ - (0.044 $\pm$ 0.008)$(V-I_c)$,\\
where $V$ and $I_c$ are the standard magnitudes; $v$ and $i$ are 
the instrumental magnitudes obtained after time and aperture corrections, and 
$X$ is the airmass. We have ignored the second-order colour correction terms 
as they are generally small in comparison to other errors present in the 
photometric data reduction. 

The standard deviations of the standardization residuals, $\Delta$, between the standard 
and transformed magnitudes and colours of the standard stars, are found to be $\Delta V = 
0.01$ mag and $\Delta (V - I_c) = 0.01$ mag. 
The photometric accuracies depend on the brightness of the stars, and the typical 
DAOPHOT errors in $V$ and $I_c$ bands at $ V\sim 18$ magnitudes are smaller than 0.01 mag. Near the 
limiting magnitude of $ V \sim 21$, which is practically the same in all the images, the 
DAOPHOT errors increase to 0.05 and 0.02 mag in the {\it V} and {\it I$_c$} bands, respectively.

\subsection {Archival Data}
\subsubsection { Near-infrared data from 2MASS}

NIR $JHK_s$ data for the stars in the BRC regions have been obtained from the Two 
Micron All Sky Survey (2MASS) Point Source Catalog (PSC) (Cutri et al. 2003). Only sources 
having uncertainty $\leq$ 0.1 mag (S/N $\geq$ 10) in all three bands were selected to ensure 
high quality data. The 2MASS $JHK_s$ data were transformed to the California Institute of Technology 
(CIT) system using the relations given in the 2MASS 
website\footnote{http://www.astro.caltech.edu/$\sim$jmc/2mass/v3/transformations/}. 

\subsubsection { Mid-infrared data from Spitzer-IRAC}

We have used the mid-infrared (MIR) data from the {\it Spitzer} archive, observed as a part of the 
``GLIMPSE360: Completing the Spitzer Galactic Plane Survey with Infrared Array Camera (IRAC)'' (PI: B. A. Whitney). 
The observations were taken during the $Spitzer$ Warm Mission, so only the data at 3.6~$\micron$ and 
4.5~$\micron$ were available. The images were taken in the High Dynamical Range mode.  
Standard Corrected Basic Calibrated Data (CBCD) products from version S18.14.0 of the {\it Spitzer} Science 
Center's IRAC pipeline were used to make the final mosaics. Both short (0.6~s) and long (12~s) integration  
CBCD frames in each channel were processed and mosaicked separately using MOPEX.  All of the mosaics were 
built at the native instrument resolution of $1^{\prime\prime}.2$/pixel. 

We used MOPEX - APEX to detect the point sources and performed the aperture photometry on the mosaicked images in each IRAC band, 
as better point-response-function images were not available for $Spitzer$ warm mission data. 
The detections were also examined visually in each band to remove non-stellar 
objects and false detections. In addition, we also included manually the point sources which were 
not detected by APEX. We supplied the list of those sources to the Apex-User list pipeline, and 
performed the aperture photometry, using an aperture radius of 2$^{\prime\prime}$.4 with the background 
estimated within a concentric sky annulus of the inner and outer radii of 2$^{\prime\prime}$.4 and 
7$^{\prime\prime}$.2,  respectively. The corresponding aperture corrections were applied as given in the IRAC data handbook V3.0 (Reach et al. 2006). 
To convert the flux densities to magnitudes, we used the zero points as  provided in the IRAC instrument handbook. 
Sources with photometric uncertainties 
less than 0.2 mag in each band were considered as good detections. We made a catalogue for each channel 
from the short and the long exposures separately 
and then looked for the closest match within $1^{\prime\prime}.2$ to compile the final catalogue of the sources detected in both IRAC bands.

\subsubsection {Wide-field Infrared Survey Explorer data}

The Wide-field Infrared Survey Explorer (WISE; Wright et al. 2010), mapping the sky in four wavebands (3.4, 4.6, 12, 
and 22~$\micron$) often referred to as W1, W2, W3, and W4, provides useful data to characterise YSOs.   
The emissions in 3.4 and 4.6~$\micron$ bands are mostly from stellar sources, and the emission in 12~$\micron$ can be
contaminated by strong polycyclic aromatic hydrocarbon (PAH) emissions at 11.3~$\micron$ and 12.7~$\micron$. 
The WISE 22 $\micron$ band traces warm dust emission from heated 
small dust grains. We have used the WISE preliminary data release catalogue\footnote{http://www.wise2.ipac.caltech.edu/docs/release/prelim/}. 
The spatial resolution of $6^{\prime\prime}$ in W1, W2 and W3, and 12$^{\prime\prime}$ in W4 may result in contamination of the 
stellar fluxes by bright extended emission or by nearby source(s), and therefore affect the identification and  
classification of stars (Megeath et al. 2004, Mercer et al. 2009). 
We rejected sources with a magnitude uncertainty $>$ 0.2 and confusion flags (cc-flags in the catalogue) that include 
any of ``D'', ``H'', ``O'' or ``P'' in the W1, W2 or W3 bands.
\subsubsection{Completeness of the data}

To evaluate the completeness of the IRAC and WISE detections, we plot the histograms of the sources for all bands. The census 
completeness limits of WISE are estimated as 11.5, 11.5, 10.5 and 6.5~mag in the W1, W2, W3 
and W4 bands, respectively. The completeness of the WISE sources is limited by the bright nebulosity and relatively
low sensitivity at the 12~$\micron$ and 22~$\micron$ bands. However, the additional data obtained from the 2MASS/IRAC photometry helps to 
cover the fainter sources. The completeness limits for the IRAC data are 15.5 and 15.0~mag in 3.6~$\micron$ 
and 4.5~$\micron$ bands, respectively. However, the identification of the IRAC/2MASS YSOs (cf. Section 4) is limited by the completeness 
limit of the K$_s$ data, which is K$_s$ $\sim$ 14.5 mag.
For BRC\,5 and BRC\,7, using the evolutionary models of Siess et al. (2000) for an age of
2~Myr and a distance of 2.1~kpc (see Section 3), the mass completeness corresponds to $\sim 0.5$~M$_\odot$.  
For BRC 39, assuming a distance of 870~$pc$ (see Section 3), we found a mass completeness limit of 0.1 M$_\odot$.

\section{Notes on individual BRC regions}

Fig.~1 shows the DSS2-R band (blue), 2MASS-$K_s$ (green) and IRAC-4.5 $\micron$ (red) colour-composite images 
of BRCs~5, 7 and 39.  A brief description of each of these BRCs is given below.

{\bf BRC 5: }
BRC\,5 is located on the west rim of the H{\sc ii} region IC\,1805 (also known as Sh2$-$190).  The bright rim 
points towards the massive stars of the cluster IC\,1805 (or Melotte\,15) located at the center of the H{\sc ii} region.  
The distance and age estimates for the cluster vary from 1.9 to 2.4~kpc and 2 to 5~Myr, respectively 
(Johnson et al. 1961, Becker 1963, Joshi \& Sagar 1983, Sung \& Lee 1995). 
In the present work, we have adopted a distance of 2.1~kpc, which we have found for the neighbouring W5 East H{\sc ii} region (Chauhan et al. 2011a). Three of the nine O stars of the cluster IC\,1805, account for 
90\% of the ionising photons of the H{\sc ii} region. These are HD\,15558 (spectral type O4\,IIIf), HD\,15570 (O4\,If), and HD\,15629 (O5\,V) 
(Massey et al. 1995). BRC\,5 is located at a projected distance $\sim 14.9$~pc from these stars.  
BRC\,5 shows signs of recent star formation as the slitless grism H$\alpha$ survey of the region by Ogura et al. (2002) revealed 
many H$\alpha$ emission stars in the region. Also, it harbours the luminous source IRAS\,02252$+$6120, with a far-infrared luminosity 
$\sim 1100$~L$_\odot$ (Sugitani et al. 2000; Lefloch, Lazareff \& Castets 1997), that powers the tiny 
jet-like HH object HH\,586 (Ogura et al. 2002). Water maser emission was detected by Xiang \& Turner (1995) towards BRC\,5, 
but was not detected later by Valdettaro et al. (2005), perhaps because of the episodic nature of the water maser emission.
Submillimeter observations indicate the presence of a dense core at the head (RA=02:29:02.2, Decl.=+61:33:33; J2000) 
of the BRC (Morgan et al. 2008). The locations of the IRAS source (cross symbol) and submillimeter core (open circle) are shown in Fig.~1a.  
The NVSS 1.4~GHz continuum emission which traces the location of the ionised boundary layer is shown with white contours. 
Since BRC\,5 satisfies all the three criteria of triggered BRC sample (cf. Morgan et al. 2009), 
it is considered as a triggered BRC.

{\bf BRC 7 : }  
BRC\,7 is a large cometary cloud located in the northern periphery of the H{\sc ii} region IC\,1805. 
It is at the projected distance of $\sim 14.6$ pc from the three most massive members of IC\,1805. 
The presence of H$\alpha$ emission stars (Ogura et al. 2002) and an IRAS point source IRAS\,02310$+$6133 with a far-infrared luminosity of $\sim 910$~L$_\odot$ near the head of the BRC\,7 (Sugitani et al. 2000) indicates ongoing star formation activity in the region. In addition to this, Wu et al. (2004) reported a CO outflow of a 
dynamical age of $\sim 5.2 \times10^5$~yr and Wouterloot, Brand \& Fiegle (1993) detected water maser 
emission in the region. However, Valdettaro et al. (2005, 2008) reported non-detection of water maser emission towards 
BRC\,7. Also, submillimeter observations (Morgan et al. 2008) of BRC\,7 at 450 and 850~$\micron$ 
revealed a dense core at the head of the cloud, where the molecular gas density is $\sim 10^4$~cm$^{-3}$. 
The locations of the IRAS source (cross symbol) and the submillimeter core (open circle) 
are shown in Fig.~1b. BRC\,7 is included in the list of triggered BRC sample of Morgan et al. (2009).
 
{\bf BRC 39 :} BRC\,39 is associated with the H{\sc ii} region IC\,1396, which is excited by the massive member(s) of the cluster 
Trumpler\,37 (Tr\,37). The most massive member and probable ionising source of the region is an O6.5\,V star, HD\,206267, 
which is at a projected distance of $\sim 13$~pc from BRC\,39. Contreras et al. (2002) 
estimated an age of about 3--5~Myr and a distance of $\sim 870$~pc for the cluster Tr\,37. IC\,1396 
has a rich population of BRCs
including BRCs 32-42 (SFO91), among which BRCs 36, 37 and 38 have been relatively well studied (see, e.g., Sicilia-Aguilar et al. 
2004, 2005; Getman et al 2007; Ikeda et al. 2008). In {\it Paper I} and {\it Paper II}, we reported quantitative evidences for 
$S^4F$ in BRC\,37 and BRC\,38, respectively.
 BRC\,39 hosts an embedded IRAS source, IRAS\,21445$+$5712, with a far-infrared 
luminosity $\sim 96$~L$_\odot$ (SFO91). It is interesting to note that the IRAS source in either BRC\,5 or BRC\,7  is located 
near the head, whereas in BRC\,39 it is located at the center of the BRC. 
With a slitless H$\alpha$ survey, Ogura et al. (2002) identified four H$\alpha$ emission stars in the region.  Moreover, 
Nakano et al. (2012) surveyed the whole IC\,1396 region, and pointed out several clusterings of H$\alpha$ emission stars  
including BRC\,39. Valdettaro et al. (2005) have reported water maser emission towards the BRC, which indicates the presence of a YSO. 
Froebrich et al. (2005) detected two bright bow shocks, HH\,865A and HH\,865B, emerging from IRAS\,21445$+$5712, pointing 
to the north-west of the BRC. The submillimeter observations by Morgan et al. (2008) at 450 and 850~$\micron$ 
revealed that BRC\,39 harbours two dense submillimeter cores, BRC39 SMM1 (RA=21:46:01.2, Decl.+57:27:42; J2000) located at the head, 
and BRC39 SMM2 (RA=21:46:06.9, Decl.+57:26:36; J2000) located in the middle of the BRC and very close to IRAS\,21445$+$5712. All of these signatures indicate ongoing star formation in the region. 
In Fig.~1c, we show the locations of the IRAS source with a cross symbol and the submillimeter cores with open circles.
BRC\,39 shows the presence of MSX emission and contains two submillimeter cores, but because the 21~cm emission feature 
was not detected above the 3$\sigma$ level, it was not considered by Morgan et al. (2009) as a triggered BRC.

\section{Identification of YSOs}\label{rd}

Since these three BRCs are located at low Galactic latitudes, the stellar contents in these regions may be significantly 
contaminated by foreground/background stars. To study the star formation scenario, it is necessary to identify the young stellar population  
associated with each region.  Here we used photometric criteria to select the YSO candidates.  
The procedure is described below.

\subsection{YSOs identified using slitless H$\alpha$ emission surveys and 2MASS photometry}

T-Tauri stars (TTSs) are low-mass YSOs, traditionally found on the basis of their H$\alpha$ or CaII emission.  
H$\alpha$ emission-line surveys are very efficient in detecting the subclass ``classical TTSs'' (CTTSs), with H$\alpha$ equivalent width 
(EW) $\ge 10$~\AA. On the other hand, weak emission (H$\alpha$ EW $< 10$~\AA) of the other subclass ``weak-line TTSs'' 
(WTTSs), usually prevents their discovery by such surveys.  In the present work, we used the catalogue of H$\alpha$ emission 
stars by Ogura et al. (2002).  In the case of BRC\,39 we supplemented the list of H$\alpha$ emission stars 
given by Nakano et al. (2012).  We searched for the additional H$\alpha$ emission stars in the catalogue of Barentsen et al. (2011) but could not find.     
Since active dMe main-sequence (MS) stars in the field also exhibit H$\alpha$ emission, some dMe stars can contaminate our young stellar 
sample. However, as the BRCs are small in sky area, the possibility of such contamination due to dMe stars is very small 
(see Ogura et al. 2002). Also, at longer 
wavelengths dMe stars do not show excess emission, so can be readily distinguished from YSOs. 

YSOs also exhibit excess emission in NIR wavelengths due to their circumstellar dust. Hence, NIR photometric surveys 
of star-forming regions are capable of detecting associated low-mass YSOs.  To identify NIR excess stars, we used 
NIR ${(J - H)/(H - K)}$ colour-colour diagrams as shown in Fig.~2. In Fig. 2, the continuous and dashed curves represent the unreddened MS and giant 
loci (Bessell $\&$ Brett 1988), respectively. 
The dotted line indicates the intrinsic colours of CTTSs (Meyer et al. 1997). 
The parallel dashed lines are the reddening vectors drawn from the tip 
(spectral type M4) of the giants (``upper reddening line''), from the base (spectral type A0) of the MS  
(``middle reddening line'') and from the tip of the intrinsic CTTSs line (``lower reddening 
line'') with crosses separated by $A_V$ $=$ 5 mag. The extinction ratios $A_J/A_V = 0.265, 
A_H/A_V = 0.155$ and $A_K/A_V=0.090$ have been adopted from Cohen et al. (1981). We classified stars according to their location in 
the colour-colour diagram (for details see {\it Paper II}). Objects falling rightward of the middle reddening vector 
and above the intrinsic CTTSs locus are considered NIR excess stars and hence probable members of the BRC aggregates. In Fig. 2, open triangles represent NIR excess stars and blue asterisk symbols represent H$\alpha$ emission stars (Ogura et al. 2002). 
In BRC\,39, the large open circles represent H$\alpha$ emission stars (Nakano et al. 2012). 
The NIR excess sources and the H$\alpha$ emission stars are included in the analyses of the present study. 

\subsection{YSOs identified using 2MASS/IRAC photometry}

MIR observations by the {\it Spitzer} Space Telescope in comparison to NIR observations provide a deeper insight into the 
YSOs which are deeply embedded in the star-forming regions. Since the 5.8~$\micron$ and 8.0~$\micron$ data are not available for our targets, 
we used the K$_s$, 3.6~$\micron$, and 4.5~$\micron$ data to identify and classify the YSOs adopting the procedure 
given by Gutermuth et al. (2009). To implement the YSO identification, we first 
generated the extinction map for the regions using the 2MASS data following the method proposed by Cambr\'{e}sy et al. (2002). In this method the region is divided into a number of cells and 
each adaptive cell contains a fixed number of stars. If $n_\ast$ is the number of stars per cell, each cell 
is defined by the closest $n_\ast$ stars to the cell center. In this way, the spatial resolution depends on the 
local stellar density, i.e., the spatial resolution is higher in a region of lower extinction.  Hence, this method 
may underestimate the extinction towards the region where dense clumps or cloud 
fragments are present. The average spatial resolution over the whole map can be changed by adjusting $n_\ast$. In our application, we have considered $n_\ast = 10$ within a typical sky area of $\sim 1$~arcmin$^2$.

We computed the colour excess $E = (H-K_s)_{\rm obs} -(H-K_s)_{\rm int}$, where $(H-K_s)_{\rm obs}$ is the observed 
median colour in a cell, and $(H-K_s)_{\rm int}$ is the intrinsic median colour estimated from the colours of supposedly 
unreddened stars. The use of median colour has the advantage of minimising the effect of foreground stars. 
To obtain the intrinsic median colour, we selected nearby field regions which were devoid of 
nebulosity or any $^{12}$CO emission features. We calculated $A_{K_s}$ for each star using the relation $A_{K_s}=1.82 E$. 
The colour excess ratios presented in Flaherty et al. (2007) have been used to compute the $(K-[3.6])_0$,  
and $([3.6]-[4.5])_0$ colours, from which Class~I and Class~II sources are selected (see Fig. 3). Although infrared wavelengths 
can penetrate dense medium to unravel the sources hidden in the dense dust layers, the background sources, especially PAH emitting galaxies and active galactic nuclei may 
also become visible in these wavelengths. To minimise these 
contaminants, we have selected objects having $[3.6]< 14.5$~mag, where the contamination from the extragalactic sources is 
expected to be less than 50\% (Fazio et al. 2004).   
Fig.~3 shows the 2MASS/IRAC $([3.6]-[4.5])_0$ vs. $(K_s -[3.6])_0$ colour-colour diagrams to identify Class~II and Class~I sources. 
Using this method we identified a total of 7 Class~I and 86 Class~II sources in BRC\,5, 1 Class~I and 55 Class~II sources 
in BRC\,7, and 3 Class~I and 25 Class~II sources in BRC\,39. 

\begin{table*}
\centering
\caption{2MASS, IRAC and WISE photometry of the identified YSOs. The complete table is available in the electronic version.}
\tiny
\hspace{-2.0cm} 
\begin{tabular}{ccccccccccccc}
\hline
Id     & R.A. & Decl. & J$\pm$eJ & H$\pm$eH & K$\pm$eK& [3.6]$\pm$ & [4.5]$\pm$ &[3.4]$\pm$ &[4.6]$\pm$&[12]$\pm$& [22]$\pm$ &comment\\
       & (J2000) &(J2000)&   &          &         &  e[3.6]    &  e[4.5]    & e[3.4]    &   e[4.6] & e[12]   & e[22]&       \\
\hline
&{\bf BRC5}&&&&&&&&&&&\\
1  & 36.97895  &61.54822 &  16.28$\pm$0.10 & 14.73$\pm$0.07 &13.80$\pm$0.05 &- & - &12.16$\pm$0.08 & 11.16$\pm$0.056 &7.74$\pm$0.28  &  -  &NIR excess\\
2  & 36.99766  &61.57376 &  14.02$\pm$0.03 & 12.78$\pm$0.03 &11.87$\pm$0.02 &- & - &10.78$\pm$0.04 & 10.18$\pm$0.032 &    -            &  -  & "\\
3  & 37.00884  &61.58534 &  16.50$\pm$0.10 & 14.94$\pm$0.06 &13.97$\pm$0.05 &13.99$\pm$0.05 & - & -               & -                 &    -            &  -  & "\\
.&.....&.....&.....&.....&.....&.....&.....&.....&.....&.....&.....&...\\
\hline
\end{tabular}
\end{table*}

\subsection{YSOs identified using WISE data}

In the absence of {\it Spitzer} longer wavelength data, we used the {\it WISE} data for further characterisation of 
the embedded YSOs. To weed out contaminants, we have exercised the approach developed by Koenig et al. (2012) which 
uses a series of colour and magnitude cuts to remove background galaxies or nebulosity blobs. 
Fig.~4 shows the colour-colour diagrams for the Class I and Class II sources identified based on the 
WISE data. We found 9 Class~I and 11 Class~II sources in BRC\,5, 4 Class~I and 20 Class~II sources in BRC\,7, and 6 Class~I 
and 19 Class~II sources in BRC\,39. Out of these 20, 24 and 25 YSOs in BRC\,5, BRC\,7 and BRC\,39, respectively, 10, 15 and 10 
YSOs are already classified as IRAC and/or 2MASS YSOs.
 
The YSOs identified and classified based on the 2MASS, IRAC and WISE photometry as well as their magnitudes in different 
bands are tabulated in Table\,2. A sample of Table\,2 is given here, whereas the complete table is available in an electronic form.

\section{Results}
\subsection{Optical properties of the selected YSOs: the colour-magnitude diagram}
The optical colour-magnitude diagram (CMD) can be used to diagnose the evolutionary status for YSOs not deeply embedded in 
the cloud.  Here we use optical $VI_c$ data to further characterise the census of YSOs.  
Fig.~5 shows the $V$ vs. $(V-I_c)$ CMDs for the YSOs identified in the present work.  
The zero-age MS as well as the pre-main-sequence (PMS) isochrones for 1 and 5 Myr for the solar metallicity by Siess et al. (2000) 
are also plotted. The isochrones are shifted for the adopted distance (see Section 3) and for the 
mean reddening of the individual BRCs, with $E(V-I_c)$ of 1.25~mag for BRC\,5, 0.93~mag for BRC\,7, and 0.50~mag 
for BRC\,39. A few stars classified as Class~I sources can also be seen in Fig. 5.  
These could be Class~II sources showing Class~I colours due to either reddening or inclination effects (Whitney et al. 2003a). The age and mass of each YSOs inferred from Fig.~5 (see {\it Paper II} for details of the methodology on the derivation) are given in Table~3. A sample of 
the Table~3 is given here and the complete table is available in an electronic form only. The ages range mostly from 0.1 to a 
few Myr. 

It is worthwhile to note that estimation of the ages of PMS stars using theoretical isochrones is prone to random errors in observations and 
systematic errors due to  different evolutionary tracks adopted (see e.g. Hillenbrand 2005). The effect of random 
errors in the determination of age and mass was estimated by propagating the random errors assuming a normal error 
distribution and using the Monte-Carlo simulations. Use of different PMS evolutionary models yields different ages 
(e.g. Sung et al. 2000) and introduces a systematic shift in age and mass determination. Here we used only the model by Siess et al. (2000) to derive the masses and the relative ages of the YSOs to avoid systematic errors. 
Binarity may be another source of error in age determination. 
An unresolved binary appears as having a higher luminosity, which consequently yields a lower age estimate on the basis of CMD.  
In the case of an equal-mass PMS binary we expect an error of $\sim50$\%--60\% in age estimation. It is, however, 
difficult to quantify the influence of binaries on the age estimation as the fraction of binaries is not known.
 
 Stellar variability arising from accretion, stellar surface activity, or circumstellar obscuration, 
may be another source of error in age determination of a YSO (Herbst et al. 1994, 1999). Moreover, the presence of a circumstellar
disk can affect its position in the CMD, leading to a wrong age (Prisinzano et al. 2011). Ages and masses derived from the PMS isochrones thus 
should be taken with caution. Also, some stars might have formed prior to the formation of the H{\sc ii} region or some may be 
background stars.  The 
location of these sources in the CMDs yield older ages. Hence, we have not included stars having ages $> 5$~Myr in further analysis. 
Here we would like to point out that we are interested mainly in the  {\it relative} ages of the aggregate members inside 
and outside bright rims.  
\begin{table*}
\centering
\caption{Magnitudes and age/mass of the YSOs in the BRC regions.The complete table is available in the electronic version. }
\begin{tabular}{cccccc}
\hline
\hline
Id     &  $V$ $\pm$ e$V$ & $I_c$ $\pm$ e$I_c$ & Mass $\pm$ error in Mass (M$_\odot$)& Age $\pm$ error Age (Myr) \\
\hline
\hline
{\bf BRC 5}&&&&&\\
   7    &   20.81  $\pm$  0.05  &  18.47  $\pm$   0.02   &0.91 $\pm$ 0.01  &    $>$ 5          \\
  10    &   22.51  $\pm$  0.13  &  18.85  $\pm$   0.07   &0.31 $\pm$ 0.02  &   1.72 $\pm$ 0.28\\
  11    &   20.08  $\pm$  0.02  &  17.64  $\pm$   0.01   &1.17 $\pm$ 0.01  &    $>$ 5         \\
... &...&...&...&...\\
\hline
\end{tabular}
\end{table*} 

\subsection{Evolutionary Status from the SED Fitting models}

To ascertain the young nature of the YSOs identified by the photometric observations, we modeled their spectral 
energy distributions (SEDs) using the fitting tools of Robitaille et al. (2006, 2007). The models are computed 
using a Monte-Carlo-based radiation transfer code (Whitney et al. 2003a, 2003b), and use combinations of the 
parameters of the central star, accreting disk, infalling envelope and bipolar cavity.  
In the present work, to constrain the physical properties of the YSOs, we fit the SEDs of only those 
sources for which we have fluxes at 12~$\micron$ or longer wavelengths. We looked for the 
counterparts of the identified YSOs in the 2MASS PSC, the WISE catalogue and optical ($V$$I_c$) data.  
For the YSOs which do not have $V$$I_c$ photometry, we resorted to the counterparts in the 3$\pi$-survey 
catalogue of Pan-STARRS1 (Kaiser et al. 2010) and used the g-, r-, i-, z-bands measurements for the SED fitting.

The SED-fitting tool deals with a single source, so if the input fluxes come from multiple sources in a beam, 
incorrect stellar age and mass are derived (Robitaille 2008). Since IRAC has a higher spatial resolution than WISE, 
we preferred the IRAC 3.6~$\micron$ and 4.5~$\micron$ fluxes wherever available. The SED-fitting tool fits 
each of the models to the data, allowing the distance and interstellar extinction to be free parameters.  
In the case of BRCs\,5 and 7, we have assumed the distance range from 1.9 to 2.4~kpc and in the case of BRC\,39 
from 0.75 to 0.87~kpc. Since we do not have spectral type information on our identified YSOs, in order to estimate 
their visual extinction (A$_V$) values, we traced the position of YSOs along the reddening vector to the intrinsic late MS locus 
or its extension in the NIR colour-colour diagram. Considering the uncertainties that might have come 
into the estimates, we used the estimated A$_V$ along with an uncertainty ($\Delta$$A_V$) of $\pm$ $2.5$~mag as an input in the SED-fitting. Class~I sources for which NIR data is not available, we allowed $A_V$ up to 25~mag. 
We further set 10\% to 30\% error in the flux estimates to allow for possible uncertainties in calibration and 
intrinsic object variability. 
In a few Class~I sources located near the head of BRCs 5, 7 and 39, we have 
also used the 450 and 850~$\micron$ fluxes from Morgan et al. (2008) for the SED fitting.  Figure~6 shows examples of the SEDs and the resulting models for Class~I and Class~II sources. 
We obtained physical parameters for all the sources adopting an approach similar to that of Robitaille et al. (2007) 
by considering those models that satisfy   
$$\chi^2 - \chi^2_{\rm min} \leq 2N_{\rm data}~,$$ 
where $\chi^2_{\rm min}$ is the
goodness-of-fit parameter for the best-fit model and $N_{\rm data}$ is the number of input observational data points.  
The parameters are obtained from the mean and the standard deviation of these models, weighted 
by $e^{({{-\chi}^2}/2)}$ of each model (see Samal et al. 2012).  Table 4 lists the Id, $\chi^2_{\rm min}$, 
stellar mass ($M_\star$), photospheric temperature ($T_\star$), stellar age ($t_\star$), mass of the disk
($M_{\rm disk}$), disk accretion rate ($\dot{M}_{\rm disk}$), foreground visual absorption ($A_V$)
 for the best fit, and the number of data points (${N}$$_{\rm data}$). 

The SED fitting reveals that the Class~I sources located near the head of BRCs 5 and 7 have relatively higher masses. In the case of BRC5$-$114, the SED fitting of the fluxes from 3.4
to 850~$\micron$ suggests a relatively massive star ($\sim$ 11 M$_\odot$). 
The Class~I source BRC7$-$74 has no 2MASS J/H counterparts but coincides 
with the submillimeter core reported by Morgan et al. (2008). The SED fitting from 2.1 to 850~$\micron$ yields a mass $\sim 6 $~M$_\odot$. 

\begin{table*}
\centering
\scriptsize
\caption{Physical parameters of the YSOs obtained from SED fits}
\begin{tabular}{cccccccccccc}
\hline\hline
 ID &${\chi}^2$$_{min}$&\multicolumn{1}{c}{$M_{\ast}$} &\multicolumn{1}{c}{$T_{\ast}$ } &\multicolumn{1}{c}{$t_{\ast}$} & \multicolumn{1}{c}{$M_{\rm disk}$ }
& \multicolumn{1}{c}{$\dot{M}_{\rm disk}$ } & \multicolumn{1}{c}{A$_V$ } &  ${N}$$_{data}$&\\

 && \multicolumn{1}{c}{($M_\odot$)}
          & \multicolumn{1}{c}{(10$^{3}$ K)}
          & \multicolumn{1}{c}{(10$^{6}$ yr)} & \multicolumn{1}{c}{($M_\odot$)} & \multicolumn{1}{c}{(10$^{-8}$ $M_\odot$/yr)} &
          \multicolumn{1}{c}{mag}&\\ 
\hline
\hline
{\bf BRC 5}&    &                &               &               &                   &                     &                &\\
1&   3.47 &  2.7 $\pm$ 1.3 & 4.4 $\pm$ 0.6 & 0.15 $\pm$ 0.13 & 0.018 $\pm$ 0.036 & 31.005 $\pm$ 29.257 & 6.5 $\pm$ 1.4 &10\\
16&  35.03 & 1.4 $\pm$ 0.9 & 4.4 $\pm$ 0.6 & 0.70 $\pm$ 0.86 & 0.003 $\pm$ 0.003 &  1.108 $\pm$  1.108 & 2.5 $\pm$ 0.4 &11\\
24&   3.45 & 1.4 $\pm$ 0.8 & 4.4 $\pm$ 0.4 & 2.26 $\pm$ 1.67 & 0.011 $\pm$ 0.013 &  4.733 $\pm$  4.397 & 4.6 $\pm$ 1.0 &8\\
27 &    0.05&  2.2 $\pm$ 0.8& 6.5 $\pm$ 2.5 &3.76 $\pm$ 3.00 &0.010 $\pm$ 0.020 &  10.93 $\pm$  11.17   &    7.4 $\pm$ 1.3 & 7 \\ 
37 &    5.96&  2.9 $\pm$ 0.7& 4.6 $\pm$ 0.2 &0.18 $\pm$ 0.06 &0.058 $\pm$ 0.009 &  27.32 $\pm$  26.71   &    1.9 $\pm$ 0.5 & 9 \\
51 &    9.49&  1.1 $\pm$ 1.2& 4.1 $\pm$ 0.8 &0.50 $\pm$ 1.62 &0.006 $\pm$ 0.015 &   3.56 $\pm$   3.55   &    2.9 $\pm$ 1.3 & 9 \\
60 &    7.58&  2.7 $\pm$ 1.8& 4.6 $\pm$ 0.6 &0.34 $\pm$ 0.42 &0.013 $\pm$ 0.038 &   7.70 $\pm$   7.69   &    3.0 $\pm$ 1.0 & 10\\
63 &   17.33&  4.9 $\pm$ 0.9& 6.1 $\pm$ 0.9 &0.34 $\pm$ 0.17 &0.044 $\pm$ 0.065 &  98.10 $\pm$  98.09   &    5.1 $\pm$ 1.1 & 9 \\
88 &    7.57&  1.2 $\pm$ 1.0& 3.9 $\pm$ 0.3 &0.07 $\pm$ 0.05 &0.011 $\pm$ 0.011 &  92.96 $\pm$  91.84   &    3.4 $\pm$ 0.9 & 7 \\
94 &    5.35&  2.2 $\pm$ 1.6& 8.3 $\pm$ 5.0 &2.18 $\pm$ 2.42 &0.006 $\pm$ 0.015 &  22.72 $\pm$  22.03   &    3.3 $\pm$ 2.2 & 9 \\
98 &    5.54&  2.1 $\pm$ 0.4& 4.6 $\pm$ 0.1 &0.33 $\pm$ 0.04 &0.005 $\pm$ 0.020 &   0.21 $\pm$   0.21   &    1.2 $\pm$ 0.2 & 11\\
102&   11.23&  0.7 $\pm$ 0.3& 3.8 $\pm$ 0.3 &0.03 $\pm$ 0.08 &0.016 $\pm$ 0.025 &  58.21 $\pm$  58.12   &    2.4 $\pm$ 0.7 & 9 \\
105&   22.44&  1.6 $\pm$ 0.3& 4.1 $\pm$ 0.1 &0.00 $\pm$ 0.00 &0.022 $\pm$ 0.023 &  64.03 $\pm$  64.03   &    8.3 $\pm$ 0.3 & 9 \\
107&    3.01&  2.3 $\pm$ 0.7& 6.3 $\pm$ 2.3 &3.52 $\pm$ 4.08 &0.008 $\pm$ 0.015 &   5.02 $\pm$   4.54   &    3.4 $\pm$ 2.0 & 8 \\
109&    3.20&  3.4 $\pm$ 1.8& 6.4 $\pm$ 5.5 &0.54 $\pm$ 1.30 &0.011 $\pm$ 0.013 & 708.80 $\pm$ 661.22   &   11.1 $\pm$ 2.5 & 5 \\
110&    2.41&  2.9 $\pm$ 1.4& 4.3 $\pm$ 1.1 &0.02 $\pm$ 0.11 &0.077 $\pm$ 0.102 & 602.45 $\pm$ 532.47   &   13.1 $\pm$ 1.9 & 7 \\
112&    0.05&  3.7 $\pm$ 1.1& 8.4 $\pm$ 4.3 &2.14 $\pm$ 2.81 &0.015 $\pm$ 0.026 &  19.89 $\pm$  22.12   &   15.0 $\pm$ 5.3 & 5 \\
113&    5.54&  1.7 $\pm$ 1.4& 4.7 $\pm$ 3.4 &0.16 $\pm$ 0.74 &0.015 $\pm$ 0.019 &1219.88 $\pm$1184.44   &   18.3 $\pm$ 3.3 & 5 \\
114&   19.08& 10.7 $\pm$ 1.4& 4.4 $\pm$ 1.4 &0.006 $\pm$ 0.03 &0.262 $\pm$ 0.295 & 646.50 $\pm$ 722.23  &   12.4 $\pm$ 5.7 & 7 \\
115&    3.77&  3.1 $\pm$ 1.9& 4.7 $\pm$ 2.2 &0.13 $\pm$ 0.39 &0.010 $\pm$ 0.019 &  37.58 $\pm$  37.26   &   11.8 $\pm$ 2.6 & 6 \\
116&   17.25&  2.6 $\pm$ 0.8& 4.7 $\pm$ 1.0 &0.40 $\pm$ 1.16 &0.011 $\pm$ 0.019 &  17.90 $\pm$  17.89   &    3.6 $\pm$ 0.6 & 11\\
117&   12.77&  5.6 $\pm$ 1.0& 4.5 $\pm$ 0.2 &0.03 $\pm$ 0.03 &0.043 $\pm$ 0.059 & 213.93 $\pm$ 213.92   &   12.1 $\pm$ 1.1 & 9 \\
118&   56.87&  3.7 $\pm$ 0.1& 6.6 $\pm$ 0.1 &1.16 $\pm$ 0.09 &0.000 $\pm$ 0.001 &   0.05 $\pm$   0.05   &   10.7 $\pm$ 0.2 & 8 \\
119&   36.65&  2.5 $\pm$ 0.7& 4.5 $\pm$ 0.2 &0.16 $\pm$ 0.06 &0.012 $\pm$ 0.017 &   9.97 $\pm$   9.97   &   9.41 $\pm$ 0.7 & 9 \\
{\bf BRC 7}&    &                &               &               &                   &                     &                &\\
 3 &  4.74 & 1.7 $\pm$ 1.0 & 4.6 $\pm$ 0.6  & 1.20 $\pm$ 1.01  & 0.021 $\pm$ 0.021  &  21.83 $\pm$ 20.80 &  6.0 $\pm$ 1.3 & 8 \\
 5 &  9.28 & 2.9 $\pm$ 0.8 & 4.8 $\pm$ 0.3  & 0.66 $\pm$ 0.43  & 0.016 $\pm$ 0.051  &   6.12 $\pm$  6.12 &  2.5 $\pm$ 0.5 & 10\\
10 & 19.33 & 2.6 $\pm$ 0.7 & 4.5 $\pm$ 0.1  & 0.21 $\pm$ 0.05  & 0.007 $\pm$ 0.012  &   3.19 $\pm$  3.19 &  7.2 $\pm$ 0.5 & 8 \\
12 &  8.58 & 2.1 $\pm$ 0.8 & 4.7 $\pm$ 0.6  & 0.73 $\pm$ 0.86  & 0.002 $\pm$ 0.008  &   1.24 $\pm$  1.24 &  1.8 $\pm$ 0.4 & 11\\
14 &  8.57 & 0.2 $\pm$ 0.1 & 3.2 $\pm$ 0.2  & 0.51 $\pm$ 0.52  & 0.004 $\pm$ 0.003  &   3.18 $\pm$  3.17 &  1.8 $\pm$ 0.7 & 10\\
15 &  0.10 & 0.9 $\pm$ 0.8 & 3.9 $\pm$ 0.4  & 3.02 $\pm$ 2.55  & 0.011 $\pm$ 0.029  &  23.63 $\pm$ 36.60 &  3.6 $\pm$ 0.8 & 7 \\
16 &  3.53 & 1.7 $\pm$ 1.0 & 4.3 $\pm$ 0.3  & 2.39 $\pm$ 2.72  & 0.010 $\pm$ 0.026  &  13.35 $\pm$ 13.15 &  2.3 $\pm$ 0.6 & 9 \\
17 & 10.09 & 1.2 $\pm$ 0.4 & 4.4 $\pm$ 0.4  & 3.01 $\pm$ 2.67  & 0.018 $\pm$ 0.012  &   8.79 $\pm$  8.77 &  3.5 $\pm$ 0.7 & 9 \\
18 & 15.67 & 0.7 $\pm$ 0.4 & 4.0 $\pm$ 0.4  & 1.99 $\pm$ 1.68  & 0.005 $\pm$ 0.003  &   0.24 $\pm$  0.24 &  1.5 $\pm$ 0.7 & 11\\
21 & 12.39 & 1.5 $\pm$ 0.3 & 4.8 $\pm$ 0.4  & 4.58 $\pm$ 3.01  & 0.004 $\pm$ 0.009  &   0.35 $\pm$  0.35 &  1.8 $\pm$ 0.5 & 9 \\
24 &  5.35 & 1.5 $\pm$ 0.4 & 4.7 $\pm$ 0.6  & 3.65 $\pm$ 2.57  & 0.009 $\pm$ 0.008  &   2.19 $\pm$  2.15 &  2.7 $\pm$ 0.7 & 9 \\
26 & 18.18 & 0.9 $\pm$ 0.3 & 4.2 $\pm$ 0.3  & 2.03 $\pm$ 1.39  & 0.001 $\pm$ 0.003  &   0.04 $\pm$  0.04 &  2.8 $\pm$ 0.4 & 8 \\
27 &  2.82 & 1.7 $\pm$ 0.4 & 5.2 $\pm$ 1.0  & 3.49 $\pm$ 3.01  & 0.007 $\pm$ 0.012  &   1.67 $\pm$  1.61 &  3.2 $\pm$ 0.8 & 8 \\
30 &  1.97 & 1.9 $\pm$ 0.6 & 4.8 $\pm$ 0.6  & 2.38 $\pm$ 2.74  & 0.004 $\pm$ 0.007  &   1.04 $\pm$  1.02 &  3.3 $\pm$ 0.7 & 8 \\
31 &  7.56 & 1.0 $\pm$ 0.8 & 4.0 $\pm$ 0.4  & 1.19 $\pm$ 1.14  & 0.006 $\pm$ 0.009  &   0.63 $\pm$  0.63 &  1.8 $\pm$ 0.9 & 8 \\
39 &  2.67 & 1.7 $\pm$ 1.1 & 4.4 $\pm$ 0.5  & 1.10 $\pm$ 2.50  & 0.007 $\pm$ 0.019  &   2.80 $\pm$  2.73 &  3.2 $\pm$ 1.1 & 8 \\
44 &  6.56 & 1.9 $\pm$ 1.4 & 4.3 $\pm$ 0.4  & 1.40 $\pm$ 1.93  & 0.021 $\pm$ 0.055  &  30.32 $\pm$ 29.81 &  3.3 $\pm$ 0.9 & 8 \\
46 &  8.87 & 2.5 $\pm$ 0.4 &10.6 $\pm$ 1.2  & 6.70 $\pm$ 1.13  & 0.001 $\pm$ 0.004  &   0.08 $\pm$  0.08 &  2.7 $\pm$ 0.3 & 9 \\
56 &  9.42 & 1.7 $\pm$ 0.9 & 4.5 $\pm$ 0.5  & 0.79 $\pm$ 1.07  & 0.026 $\pm$ 0.028  &   8.58 $\pm$  8.55 &  2.1 $\pm$ 0.7 & 9 \\
59 & 32.94 & 3.0 $\pm$ 0.2 & 7.5 $\pm$ 0.6  & 2.35 $\pm$ 0.48  & 0.001 $\pm$ 0.005  &   1.61 $\pm$  1.61 &  8.1 $\pm$ 0.3 & 9 \\
64 &  9.85 & 2.0 $\pm$ 0.4 & 7.5 $\pm$ 2.8  & 4.97 $\pm$ 2.43  & 0.007 $\pm$ 0.012  &   6.40 $\pm$  6.39 &  3.4 $\pm$ 1.2 & 11\\
65 &  4.54 & 3.6 $\pm$ 0.2 & 6.5 $\pm$ 1.1  & 1.16 $\pm$ 0.34  & 0.000 $\pm$ 0.000  &   0.00 $\pm$  0.00 &  1.0 $\pm$ 0.0 & 7 \\
66 &  5.79 & 1.3 $\pm$ 0.2 & 4.6 $\pm$ 0.4  & 2.97 $\pm$ 2.44  & 0.006 $\pm$ 0.006  &   8.78 $\pm$  8.60 &  5.1 $\pm$ 0.5 & 10\\
67 & 21.30 & 2.8 $\pm$ 0.5 & 7.0 $\pm$ 3.6  & 2.32 $\pm$ 3.23  & 0.016 $\pm$ 0.026  &  18.72 $\pm$ 18.72 &  9.7 $\pm$ 1.9 & 8 \\
68 & 39.32 & 1.9 $\pm$ 1.2 & 4.0 $\pm$ 0.5  & 0.14 $\pm$ 0.14  & 0.012 $\pm$ 0.033  &   9.26 $\pm$  9.26 &  5.4 $\pm$ 1.2 & 8 \\
69 &  0.88 & 2.0 $\pm$ 0.4 & 7.4 $\pm$ 2.1  & 5.60 $\pm$ 2.97  & 0.004 $\pm$ 0.009  &   1.92 $\pm$  1.99 &  4.8 $\pm$ 0.9 & 5 \\
70 &  0.66 & 2.1 $\pm$ 0.9 & 5.6 $\pm$ 2.1  & 2.52 $\pm$ 3.57  & 0.032 $\pm$ 0.048  &  23.62 $\pm$ 21.95 &  4.7 $\pm$ 1.9 & 6 \\
71 &  4.49 & 1.4 $\pm$ 0.5 & 4.4 $\pm$ 0.2  & 0.51 $\pm$ 0.61  & 0.007 $\pm$ 0.015  &   3.02 $\pm$  2.85 &  2.5 $\pm$ 0.4 & 6 \\
72 &  0.64 & 2.0 $\pm$ 0.4 & 8.8 $\pm$ 1.8  & 7.23 $\pm$ 2.52  & 0.002 $\pm$ 0.003  &   2.27 $\pm$  2.14 &  8.2 $\pm$ 1.4 & 5 \\
73 & 12.62 & 1.8 $\pm$ 0.2 & 5.6 $\pm$ 1.3  & 5.57 $\pm$ 1.70  & 0.005 $\pm$ 0.006  &   0.34 $\pm$  0.34 &  4.5 $\pm$ 0.8 & 8 \\
74 & 59.62 & 6.0 $\pm$ 0.3 & 4.2 $\pm$ 0.1  & 0.01 $\pm$ 0.01  & 0.042 $\pm$ 0.053  &    370.60 $\pm$ 235.50  & 13.0 $\pm$ 0.9 & 7 \\
76 &  2.89 & 1.1 $\pm$ 1.2 & 4.1 $\pm$ 2.1  & 0.11 $\pm$ 0.52  & 0.021 $\pm$ 0.035  & 365.24 $\pm$ 327.78& 12.5 $\pm$ 4.0 & 5 \\ 
{\bf BRC 39}&    &                &               &               &                   &                     &                &\\
 5 &11.12  &0.5 $\pm$ 0.2 &3.8 $\pm$ 0.2  &6.17 $\pm$ 0.60  &0.000 $\pm$ 0.001  &  2.04 $\pm$   2.04      &  3.8  $\pm$     0.1 & 10\\
17 & 2.00  &0.3 $\pm$ 0.1 &3.5 $\pm$ 0.1  &5.32 $\pm$ 1.69  &0.001 $\pm$ 0.001  &  0.09 $\pm$   0.08      &  1.4  $\pm$     0.3 &  8\\
22 &13.07  &0.1 $\pm$ 0.0 &2.9 $\pm$ 0.0  &1.23 $\pm$ 0.66  &0.002 $\pm$ 0.002  &  0.23 $\pm$   0.23      &  1.4  $\pm$     0.1 &  9\\
23 &22.17  &0.3 $\pm$ 0.0 &3.5 $\pm$ 0.1  &0.26 $\pm$ 0.32  &0.000 $\pm$ 0.001  &  0.06 $\pm$   0.06      &  2.9  $\pm$     0.2 &  9\\
26 & 2.14  &2.1 $\pm$ 0.8 &4.7 $\pm$ 0.4  &0.64 $\pm$ 0.56  &0.005 $\pm$ 0.018  &  0.05 $\pm$   0.05      &  9.9  $\pm$     0.9 &  8\\
27 &22.55  &0.8 $\pm$ 0.4 &3.9 $\pm$ 0.4  &0.07 $\pm$ 0.46  &0.003 $\pm$ 0.005  & 42.99 $\pm$  42.99      &  9.4  $\pm$     0.3 &  9\\
28 & 3.51  &0.1 $\pm$ 0.0 &2.9 $\pm$ 0.1  &1.46 $\pm$ 0.91  &0.000 $\pm$ 0.000  &  0.01 $\pm$   0.01      &  1.3  $\pm$     0.1 & 10\\
31 & 8.66  &0.2 $\pm$ 0.0 &3.3 $\pm$ 0.1  &0.07 $\pm$ 0.01  &0.020 $\pm$ 0.001  & 65.71 $\pm$  65.07      &  3.8  $\pm$     0.4 & 11\\
33 & 6.59  &0.1 $\pm$ 0.0 &3.1 $\pm$ 0.1  &1.04 $\pm$ 0.48  &0.005 $\pm$ 0.003  &  1.00 $\pm$   0.98      &  1.8  $\pm$     0.3 & 11\\
39 &19.15  &0.1 $\pm$ 0.0 &3.1 $\pm$ 0.1  &0.27 $\pm$ 0.13  &0.006 $\pm$ 0.001  &  4.62 $\pm$   4.62      &  3.8  $\pm$     0.8 &  9\\
40 & 8.77  &1.0 $\pm$ 0.7 &4.3 $\pm$ 1.5  &0.70 $\pm$ 1.71  &0.036 $\pm$ 0.028  & 72.26 $\pm$  71.57      &  2.9  $\pm$     1.1 &  9\\
41 &15.88  &1.9 $\pm$ 0.1 &4.1 $\pm$ 0.1  &0.003 $\pm$ 0.001  &0.052 $\pm$ 0.004  &641.16 $\pm$ 640.94      &  7.3  $\pm$     0.2 & 9\\
42 & 1.83  &2.3 $\pm$ 0.5 &9.2 $\pm$ 1.7  &6.38 $\pm$ 1.81  &0.010 $\pm$ 0.016  &  2.02 $\pm$   2.01      &  3.5  $\pm$     0.5 &  9\\
\hline
\end{tabular}
 \end{table*}
\begin{table*}
\scriptsize
\centering
\begin{tabular}{cccccccccccc}
\bf{Table 4. Cont.}&&&&&&&&&&&\\
\hline\hline
46 & 0.26  &0.3 $\pm$ 0.3 &3.2 $\pm$ 0.4  &0.72 $\pm$ 1.67  &0.004 $\pm$ 0.005  &  3.05 $\pm$  2.71  & 15.26 $\pm$ 4.98 &  5\\
47 &13.24  &3.4 $\pm$ 1.6 &6.3 $\pm$ 3.4  &1.01 $\pm$ 1.76  &0.016 $\pm$ 0.031  &18.84  $\pm$ 18.84  &  3.81 $\pm$ 1.93 & 7\\
48 & 5.28  &1.5 $\pm$ 0.2 &4.5 $\pm$ 0.2  &1.38 $\pm$ 1.09  &0.007 $\pm$ 0.016  &  1.23 $\pm$  1.22  &  1.93 $\pm$ 0.66 & 10\\
50 & 0.82  &1.2 $\pm$ 0.5 &4.3 $\pm$ 0.4  &1.24 $\pm$ 1.01  &0.002 $\pm$ 0.005  &  0.87 $\pm$  1.78  &  14.8 $\pm$ 1.02 &  6\\
51 & 3.19  &0.3 $\pm$ 0.1 &3.2 $\pm$ 0.3  &3.83 $\pm$ 3.99  &0.003 $\pm$ 0.002  &  3.24 $\pm$  2.91  &  1.84 $\pm$ 0.60 &  7\\
52 &34.64  &2.7 $\pm$ 0.1 &4.9 $\pm$ 0.1  &1.18 $\pm$ 0.09  &0.000 $\pm$ 0.000  &  0.01 $\pm$  0.01  &  2.82 $\pm$ 0.30 & 10\\
53 & 8.08  &0.8 $\pm$ 0.3 &4.0 $\pm$ 0.6  &1.60 $\pm$ 1.43  &0.002 $\pm$ 0.005  &  9.41 $\pm$  9.31  &  2.93 $\pm$ 0.56 & 11\\
54 &20.37  &2.5 $\pm$ 0.6 &5.3 $\pm$ 0.6  &2.02 $\pm$ 1.14  &0.001 $\pm$ 0.008  &  0.49 $\pm$  0.49  &  2.81 $\pm$ 0.78 & 11\\
55 &30.21  &2.6 $\pm$ 0.7 &5.0 $\pm$ 0.5  &1.22 $\pm$ 0.44  &0.030 $\pm$ 0.010  & 32.53 $\pm$ 32.53  &  5.25 $\pm$ 0.52 & 11\\
56 &16.16  &1.9 $\pm$ 0.2 &4.8 $\pm$ 0.3  &2.79 $\pm$ 1.95  &0.005 $\pm$ 0.009  &  6.07 $\pm$  6.07  &  2.45 $\pm$ 0.32 & 10\\
57 &16.87  &0.2 $\pm$ 0.0 &3.0 $\pm$ 0.1  &0.21 $\pm$ 0.20  &0.007 $\pm$ 0.004  &  5.63 $\pm$  5.63  &  2.78 $\pm$ 0.28 & 10\\
58 & 0.81  &0.7 $\pm$ 0.5 &3.8 $\pm$ 0.8  &1.41 $\pm$ 1.50  &0.004 $\pm$ 0.008  &  1.06 $\pm$  0.99  &  9.10 $\pm$ 1.90 &  6\\
59 &11.64  &0.2 $\pm$ 0.1 &3.1 $\pm$ 0.2  &1.08 $\pm$ 1.33  &0.003 $\pm$ 0.001  &  3.35 $\pm$  3.34  &  2.76 $\pm$ 0.42 & 10\\
\hline
\end{tabular}
 \end{table*}
\begin{table*}
\scriptsize
\begin{tabular}{cccccccccccc}
\hline
\hline
\end{tabular}
 \end{table*}

\section{Discussion}

\subsection{Mid/Far Infrared View of BRCs}

Figure\,7 shows spatial distribution of the YSOs obtained from the 2MASS, IRAC and WISE data, overplotted on the three-colour composite images (WISE 3.4, 12 and 22~$\micron$) of BRCs~5, 7 and 39. In Fig.~7, open triangles represent NIR excess sources from 2MASS, squares and diamonds represent the IRAC Class~I 
and Class~II sources, respectively. Class~I and Class~II sources selected based on the WISE colours are
 shown as crosses and open circles, respectively. Also plotted are the contours showing the data from the Bolocam Galactic Plane Survey (BGPS; Aguirre et al. 2011), a 1.1~mm continuum 
survey using the Bolocam camera on the Caltech Submillimeter Observatory (CSO). The BGPS images 
and source catalogues are available through http://irsa.ipac.caltech.edu. 
  
 The millimeter continuum traces emission from a dense dust core, which may eventually form a star or a stellar 
system.  In Fig.~7a, we notice that the 1.1~mm emission encloses the Class I sources at the head of BRC\,5 and BRC\,7.
Slightly to the south-west of BRC\,5, there is a filamentary distribution of a few red (mainly Class I) sources enclosed by the 1.1~mm 
emission contours. 
In the case of BRC\,39, 
the position of the 1.1~mm core coincides with the BRC\,39~SMM1 (marked in Fig.~1c). 

 The spatial distribution of the YSOs in the three BRC regions studied here  
(see Fig.~7) reveals that the Class~II sources are located mostly outside the bright rims, while the Class~I sources 
are generally located within the bright rims. In case of BRC\,7, we can see an aligned distribution of YSOs (see Fig.~7b) 
towards the direction of ionising sources of the region. 
Similarly, the spatial distribution of YSOs towards BRC\,39 (shown in Figure~7c) also indicates an increased number 
near the head of the globule. 
Additionally, we also notice the intricate structures with a few globules smaller in size than BRC in the region.  
A few of these globules are marked in Figure~7c.  
These small globules are pointing towards the massive stars of Tr\,37 and generally contain stars at their tips, 
whose IRAC/WISE colours suggest Class I/Class II nature.  
These are morphologically similar to the elephant trunk-like structures (ETLSs) found in the H{\sc ii} region W5\,E (Chauhan et al. 2011b).

\subsection{ Small-Scale Sequential Star Formation ({\it $S^4$F})}
\begin{table}
\caption{Mean age of the YSOs inside/on and outside a bright rim}
\begin{tabular}{|p{.4in}|p{1.2in}|p{1.2in}|}
\hline

Region &Inside/on bright rim & Outside bright rim\\
       &Mean age (Myr)        & Mean age (Myr)\\
       &(No. of stars)       & (No. of stars)\\
\hline
BRC 5   & 1.76 $\pm$ 1.17 (19) & 2.06 $\pm$ 1.24 (17)\\
BRC 7   & 0.92 $\pm$ 0.74 (12) & 1.35 $\pm$ 0.77 (26)\\ 
BRC 39  & 2.01 $\pm$ 1.65  (5) & 2.51 $\pm$ 1.31 (15)\\
\hline
\end{tabular}
\end{table}

 In {\it Paper I} and {\it Paper II}, we have quantitatively verified the $S{^4}F$ hypothesis by using {\it BVI$_{c}$} photometry of a few BRCs. 
In the present study we follow the same approach by classifying the YSO population (see Section 4) associated with a BRC into two groups:~those lying on/inside 
versus outside of a rim. The mean ages calculated for the two groups (from Table~3), given in Table~5, show that the YSOs lying on/inside the rim 
are younger than those located outside. For example, in the case of BRC\,7, the stellar ages 
decrease from 4~Myr for the central cluster, to 1.4~Myr outside the rim, and further to $\sim 0.9$~Myr 
inside/on the rim. These results are similar to those obtained in {\it Paper I} and {\it II}, so reinforcing the $S{^4}F$ hypothesis.   
Although BRC\,39 is not included as a triggered star-forming candidate in Morgan et al. (2009), our analysis shows an age gradient with the location of YSOs  
relative to the bright rim, so BRC\,39 should also manifest triggered star formation.  

Although the gradient of the mean age is clear in each BRC, the scatter of age is large. 
The projection effect may cause such a scatter, e.g., relatively older stars located in the outer layers of a 
cloud may be projected inside or on the rim. The present data does not allow us to correct for the effect, but 
given equal probability of projection, it should not affect our conclusion.
Alternative possibilities for the scatter include errors in photometry or in extinction correction, stellar variability, binarity, and proper  
motions of stars etc. As discussed in {\it Paper I}, photometric errors, variability and extinction correction do not seem to be the main reason for the 
scatter, but proper motion could be a probable reason for the scatter in the mean ages.  

In Table~6, we have tabulated various information regarding the BRCs studied in the present work, from our earlier work, and from Getman et al. (2012). 
Table~6 includes the projected distances of the BRCs from the corresponding ionising star(s), the possible ages of the ionising star clusters, 
the relative sizes of the BRCs and the ages of in-cloud as well as off-cloud YSOs.  
The distances from the ionising sources and the relative sizes of the BRCs are taken from SFO91.  For BRC\,36, we have adopted 
the in-cloud and off-cloud YSO ages from 
Getman et al. (2012) whereas for the rest of the BRCs, the mean ages are taken from our earlier ($Paper I$, $Paper II$ and Chauhan et al. 2011a) and present 
works. In every case the mean age of the YSOs in the vicinity of a BRC appears to be younger than the of the cluster containing ionising star(s), supporting the occurrence of the triggering processes in the region.
\begin{table*}
\caption{A comparison of different BRC regions}
\begin{tabular}{|p{.8in}|p{1.3in}|p{1.0in}|p{0.5in}|p{0.5in}|p{1.0in}|p{1.0in}|}
\hline
Region &Projected distance from&Age of the ionising &length  &width& inside/on age&Outside age\\
       &the ionising star(s)(pc) &cluster (Myr)&(pc) &(pc)& (Myr)&(Myr)\\

\hline
BRC 2  &    16.9&  $\sim$2$^a$ &  0.46& 0.86   &  0.90 $\pm$ 1.00 &    -          \\
BRC 5  &   14.9 &  3-5  &  1.33& 1.40   &  1.76 $\pm$ 1.17 & 2.06 $\pm$ 1.24 \\
BRC 7  &   14.6 &  3-5&  2.11& 3.33   &  0.91 $\pm$ 0.71 & 1.38 $\pm$ 0.77\\
BRC 11 &   11.9 &  3-5$^b$&  0.53& 1.05   &  0.86 $\pm$ 0.71 & 1.26 $\pm$ 0.66 \\
BRC 12&     3.9 &  3-5$^b$&  1.23& 1.05   &  0.30 $\pm$ 0.17 & 2.51 $\pm$ 2.43\\
BRC 13 &    7.0 &  2-5$^{b,c}$&  2.04& 1.93   &  1.61 $\pm$ 1.41 & 2.44 $\pm$ 1.37 \\
BRC 14 &    8.6 &  2-5$^{b,c}$&  0.98& 2.91   &  1.02 $\pm$ 0.67 & 2.32 $\pm$ 1.22 \\
BRC 36&     4.9 &  $\sim$4&  -   & 0.97   &   $\sim$ 1 & 2-3\\
BRC 37&    11.9 &  $\sim$4&  0.65& 0.19   &  0.92 $\pm$ 0.63 & 1.93 $\pm$ 2.22\\
BRC 38 &   10.4 &  $\sim$4&  0.76& 1.11   &  2.00 $\pm$ 0.90 & 2.70 $\pm$ 0.90\\
BRC39&   12.6   & $\sim$4& 1.01&  0.60 &      2.01 $\pm$ 1.65&2.67 $\pm$ 1.20\\
\hline
\end{tabular}
$^a$ Pandey et al. 2008\\
$^b$ Karr \& Martin 2003\\
$^c$ Chauhan et al. 2011a\\
\end{table*}

\subsection{Possible sources of UV radiation}

BRCs\,5 and 7 both host small clusters/aggregates of YSOs.  The clustering in BRC\,7 is prominent, visible in 2MASS $K_s$ as well 
as in IRAC and WISE images. Its formation may be the result of the 
implosion of the original cloud which has now receded slowly due to the photoevaporation 
of the molecular material. The slightly elongated morphology of the cluster further strengthens 
the preposition.  

We examined the elongation of the cluster and the distribution of NIR excess sources and of Class~I/II sources obtained 
from the IRAC/WISE photometry as well as the symmetry axis of the cloud. In the case of BRC\,7 the direction from the mean position of the three most 
massive stars in IC\,1805 to the IRAS source (as marked in Fig.~7b) is nearly parallel to the elongation direction of 
the BRC, suggesting that the cloud has been strongly affected by the UV radiation from the luminous stars. 
However, the distribution of the Class~II sources has a slightly different elongation direction from that to the massive stars. Instead, 
the elongated distribution points to an O6 V star (BD\,$+$60\,497). This suggests that initially the influence of BD\,$+$60\,497 may have been dominant 
in forming and shaping the cluster and the present main exciting stars may have been born later than BD\,$+$60\,497. 
Also the presence of millimeter cores and 21-cm continuum emission to the west of the cloud (see Fig.~1b) indicates much exposure to 
the UV radiation on the western side. The morphology of the tip of the rim also supports the idea. 

Since BRC\,5 is also located in the same H{\sc ii} region and its evolution must have been subjected to the same ionising stars, 
we also examined the possible role of BD\,$+60$\,497 on the evolution of the cloud.  Lefloch, Lazareff \& Castet (1997), using molecular line and continuum 
observations, compared the distribution of ionised gas along the cloud rim and the general symmetry of the cloud. They
suggested that the cloud acquired its shape in a previous episode of photoevaporation by now bygone stars and that it
is presently undergoing recompression due to the radiation from the three most massive stars of the cluster located towards the south direction.  
BD\,$+$60\,497 is nearly aligned with the apex of BRC\,5, so appears to be the most likely source responsible for the current morphology of BRC\,5.  
However, an O9.5 V star adjacent to the BRC (shown with small diamond symbol in Fig.~7a) may also have some influence.

In the case of BRC\,39, the DSS2-R band image shows that neither the symmetry axis of the BRC, nor the curvature of the rim points towards the most massive star HD\,206267 (see Figure~7c). 
To explore the possibility of alternative ionising star(s) which may influence the geometry and evolution of the cloud, we searched for other massive stars 
(earlier than B2) in the region using the SIMBAD database. We found HD\,206773, a B0\,V star and a member of Tr\,37, to be the closest massive star to BRC\,39 and 
somewhat aligned with the apex of the cloud. The projected distance of the apex of the BRC from HD\,206773 is nearly half of that from HD\,206267, but 
the contribution of the UV flux from HD\,206773 is only 21\% (neglecting the projection effect) 
of that from HD\,206267.

\subsection{Indication of Global Triggered Star Formation}                                        

BRCs are considered to be remnants from the dense parts (cores/clumps) of an inhomogeneous giant molecular cloud. 
If the original molecular cloud was big enough, it may have caused a series of triggered star formation events, leaving OB 
association/stellar aggregates of different generation. The spatial distribution and ages of such aggregates could be used to 
probe the star formation history of the giant molecular cloud. 
In the present section, we discuss the global star formation in the H{\sc ii} regions associated 
with the studied BRCs.

Oey et al. (2005) studied the hierarchical triggering by the superbubble in the W3/W4 region.  The cluster IC\,1805 (age $\sim 2$--5~Myr) is located near the center of W4, and the nearby cluster IC\,1795 in W3 has a similar age of $\sim 3$--5~Myr, while the 
embedded populations in W3-North, W3-Main and W3-OH are younger ($\sim 10^5$--$10^6$~yr). On the other hand 
the age of the shell structure of the superbubble is estimated to be $\sim 6$--10~Myr (Dennison et al. 2007).
 So the shell structure should have been formed by the oldest generation of stars in the W4 region. We suspect
 that those stars consequently gave birth to IC 1795 and IC 1805. Then IC\,1795 in turn triggered the formation of currently embedded stars in W3-North, W3-Main and W3-OH. Similarly, the IC\,1805 cluster is currently triggering a younger population at its periphery, in BRC\,5 and BRC\,7. 

Patel et al. (1995) studied IC\,1396 using $^{12}$CO and $^{13}$CO observations. They pointed out that 
a prominent group of BRCs in the region forms a ring around the cluster Tr\,37. The ring is expanding at a speed of $\sim 5$~km~s$^{-1}$. The dynamical age of the system ($\sim2$--3~Myr) 
matches very well with the age of BRC\,39 found in the present work. 
 Sicilia-Aguilar et al. (2004, 2005) estimated the ages of TTSs in the cluster Tr\,37 as $\sim$ 4 Myr, and an younger TTSs  
population ($\sim 1$~Myr) associated with BRC\,36. Recently, Barentsen et al. (2011) and Getman et al. (2012) found that for BRC\,36, the stellar age increases from 
$\sim 1$~Myr inside the cloud, to $\sim 2$--3~Myr in front of the cloud, to $\sim 4$~Myr towards Tr\,37.  
In the present work, we found a similar trend for BRC\,39, where the mean ages of the YSOs increases from inside/on the bright rim ($\sim 2$~Myr), to 
outside bright rim ($\sim 2.5$~Myr), and further increase towards Tr\,37 ($\sim 4$~Myr).

\section{Conclusions}

On the basis of optical and IR observations to characterise the young stellar aggregates of BRC\,5, BRC\,7, and BRC\,39, we confirm the significant role played by massive stars on the star formation activity in neighboring clouds.  
In each bright-rimmed cloud studied here, there is a spatial orientation between the molecular cloud with an ionised rim and one or a group of massive stars. Moreover, 
the mean ages of the YSOs inside/on the rim are found to be younger than that of the stars outside, confirming the 
scenario of S$^4$F.  Our study provides further manifestation of massive stars triggering birth of next-generation 
star aggregates/groups.  In BRC\,7, we found a group of Class II sources in front of the rim, which may be the semi-final product of the implosion of a fragment of the originally massive cloud, indicating self-propagating star formation. 
 
\section{ACKNOWLEDGMENTS}
We are extremely thankful to the anonymous referee for his/her patient and critical comments which significantly improved the contents of
the paper significantly. The work at NCU is financially supported by the National Science Council through 
the grant no.~102-2119-M-008-001. NP also acknowledges financial support from the Department of Science \& Technology,
INDIA, through INSPIRE faculty award~IFA-PH-36. This publication makes use of data from the Two Micron All Sky Survey (a 
joint project of the University of Massachusetts and the Infrared Processing 
and Analysis Center/ California Institute of Technology, funded by the 
National Aeronautics and Space Administration and the National Science 
Foundation), archival data obtained with the {\it Spitzer Space Telescope}
(operated by the Jet Propulsion Laboratory, California Institute 
of Technology, under contract with the NASA) and archival data obtained from the Wide Infrared 
Survey Explorer (WISE).

\section*{References}
Aguirre J.E., Ginsburg A.G., Dunham M.K., Drosback M.M., Bally J. et al. 2011, ApJS, 192, 4\\
Barentsen G., Vink J.S., Drew J.E., Greimel R., Wright N.J. et al. 2011, MNRAS, 415, 103\\
Becker W. 1963, ZA, 57, 117\\
Bertoldi F., 1989, ApJ, 346, 735\\
Bessell M.S., \& Brett J.M. 1988, PASP, 100, 1134\\
Cambr{\'e}sy L., Beichman C.A., Jarrett T.H., Cutri R.M 2002, AJ, 123, 2559\\
Chauhan N., Pandey A.K., Ogura K., Ojha D.K., Bhatt B.C., Ghosh S.K., Rawat P.S. 2009, MNRAS, 396, 964 ({\it Paper II})\\
Chauhan N., Pandey A.K., Ogura K., Jose J., Ojha D.K. et al. 2011a, MNRAS, 415, 1202\\
Chauhan N., Ogura K., Pandey A.K., Samal M.R., \& Bhatt B.C. 2011b, PASJ, 63, 795\\
Cohen J.G., Frogel J.A., Persson S.E., \& Ellias J.H. 1981, ApJ, 249, 481\\
Contreras M.E., Sicilia-Aguilar A., Muzerolle J., Calvet N., Berlind P., Hartmann L. 2002, ApJ, 124, 1585\\
Cutri R.M., Skrutskie M.F., Van Dyk S. et al. 2003, Vizier Online Data Catalog, 2246, O\\
Dennison B., Topasna G.A. \& Simonetti J.H. 1997, ApJ, 474, L31\\
Elmegreen B.G., \& Lada C.J. 1977, ApJ, 214, 725\\
Fazio G.G., Hora J.L., Allen L.E., Ashby M.L.N., Barmby P. et al. 2004, ApJS, 154, 39\\
Flaherty K.M., Pipher J.L., Megeath S.T., Winston E.M., Gutermuth R.A. et al. 2007, ApJ, 663,1069\\
Froebrich D., Scholz A., Eisl{\"o}ffel J., \& Murphy C.C. 2005, A\&A, 432, 575\\
Getman K.V., Feigelson E.D., Garmire G., Broos P., \& Wang J. 2007, ApJ, 654, 316\\
Getman K.V., Feigelson E.D., Sicilia-Aguilar A., Broos P.S., Kuhn M.A. 2012, MNRAS, 426, 2917\\
Gutermuth R.A., Megeath S.T., Myers P.C., Allen L.E., Pipher J.L., Fazio G.G. 2009, ApJS, 184,18\\
Herbst W., Herbst D.K., Grossman E.J. \& Weinstein D. 1994, AJ, 108, 1906\\
Herbst W. \& Shevchenko V.S. 1999, AJ, 118, 1043\\
Hillenbrand L.A. 2005, A Decade of Discovery: Planets Around Other Stars" STScI Symposium Series 19, ed. M. Livio, astro-ph/0511083\\
Ikeda H., Sugitani, K., Watanabe M. et al. 2008, AJ, 135, 2323\\
Johnson H.L., Hoag A.A., Iriarte B., Mitchell R.I., Hallam K.L. 1961, LowOB, 5, 133\\
Jose J., Pandey A.K., Ogura K., Ojha D.K., Bhatt B.C. et al. 2012, MNRAS, 411, 2530\\
Joshi U.C. \& Sagar R. 1983, JRASC, 77, 40\\
Karr J.L. \& Martin P.G. 2003, ApJ, 595, 900\\
Kaiser N., Burgett W., Chambers K., Denneau L., Heasley J. et al. 2010, 
in Society of Photo-Optical Instrumentation Engineers (SPIE) Conference Series, volume 7733 \\
Koenig X.P., Leisawitz D.T., Benford D.J., Rebull L.M., Padgett D.L., Assef R.J. 2012, ApJ, 744, 130\\
Landolt A.U. 1992, AJ, 104, 340\\
Lee H.T., Chen W.P., Zhang Z.W., Hu J.Y. 2005, ApJ, 624, 808\\
Lefloch B., Lazareff B. 1995, A\&A, 301, 522\\
Lefloch B., Lazareff B., Castets A. 1997, A\&A, 324, 249\\
Massey P., Johnson K. E., Degioia-Eastwood K. 1995, ApJ, 454, 151 \\
Megeath S.T., Allen L.E., Gutermuth R.A., Pipher J.L., Myers P.C. et al. 2004, ApJS, 154, 367\\
Mercer E.P., Miller J. M., Calvet N., Hartmann L., Hernandez J. et al. 2009, AJ, 138, 7\\
Meyer M., Calvet N., \& Hillenbrand, L.A. 1997, AJ, 114, 288\\
Morgan L.K., Thompson M.A., Urquhart J.S., White G.J. 2008, A\&A, 477, 557\\
Morgan L.K., Urquhart J.S., Thompson M.A. 2009, MNRAS, 400, 1726\\
Nakano M., Sugitani K., Watanabe M., Fukuda N., Ishihara D., Ueno M. 2012, AJ, 143, 61\\
Ogura K., Sugitani K., Pickles A. 2002, AJ, 123, 2597\\
Ogura K., Chauhan N., Pandey A.K., Bhatt B.C., Ojha D.K., Itoh Y. 2007, PASJ, 59, 199 ({\it Paper I})\\
Oey M.S., Watson A.M., Kern K., Walth G.L. 2005, AJ, 129, 393\\
Pandey A.K., Sharma S., Ogura K., Ojha D.K., Chen W.P., Bhatt B.C., Ghosh S.K. 2008, MNRAS, 383, 1241\\
Patel N.A., Goldsmith P.F., Snell R.L., Hezel T., Xie T. 1995, ApJ, 447, 721\\
Prisinzano L., Sanz-Forcada J., Micela G., Caramazza M., Guarcello M.G. et al. 2011, A\&A, 527, 77\\
Reach, W. et al. 2006, Infrared Array Camera Data Handbook, version 3.0, Spitzer Science Centre, California Institute of Technology, Pasadena, California 91125 USA\\
Robitaille T.P., Whitney B.A., Indebetouw R., Wood K., Denzmore P. 2006, ApJS, 167, 256\\
Robitaille T.P., Whitney B.A., Indebetouw R., \& Wood K. 2007, ApJS, 169, 328\\
Robitaille T.P. 2008, ASPC, 387, 290\\
Samal M.R., Pandey A.K., Ojha D.K., Chauhan N., Jose, J. et al. 2012, ApJ, 755, 20\\
Sicilia-Aguilar A., Hartmann L.W., Briceno C., Muzerolle J., Calvet N. 2004, AJ, 128, 805\\
Sicilia-Aguilar A., Hartmann L.W., Harnandez J., Briceno C., Calvet N. 2005, AJ, 130, 188\\
Siess L., Dufour E., Forestini M. 2000, A\&A, 358, 593\\
Stetson P.B. 1987, PASP, 99, 191\\
Stetson P.B. 1992, ASPC, 25, 297\\
Sugitani K., Fukui Y., Ogura K. 1991, ApJS, 77, 59 (SFO91)\\
Sugitani K., Ogura K. 1994, ApJS, 92, 163\\
Sugitani K., Tamura M., Ogura K. 1995, ApJ, 455, L39\\
Sugitani K., Matsuo H., Nakano M., Tamura M., Ogura K. 2000, AJ, 119, 323\\
Sung H. \& Lee S.-W. 1995, JKAS, 28, 119\\
Sung H., Chun M. Y., Bessel M.S. 2000, AJ, 120, 333\\
Valdettaro R., Migenes V., Trinidad M. A., Brand J., Palla F. 2008, ApJ, 675, 1352\\
Valdettaro R., Palla F., Brand J., \& Cesaroni R. 2005, A\&A, 443, 535\\
Whitney B.A., Wood K., Bjorkman J.E., \& Cohen M. 2003a, ApJ, 598, 1079\\
Whitney B.A., Wood K., Bjorkman J.E., \& Wolff M.J. 2003b, ApJ, 591, 1049\\
Wouterloot J.G.A., Brand J. \& Fiegle K. 1993, A\&AS, 98, 589\\
Wright E.L., Eisenhardt P.R.M., Mainzer A.K., Ressler M.E., Cutri R.M., et al. 2010, AJ, 140, 1868\\
Wu Y., Wei Y., Zhao M., Shi Y., Yu W., Qin S., Huang M. 2004, A\&A, 426, 503\\
Xiang D. \& Turner B.E. 1995, ApJS, 99, 121\\
\bsp
\label{lastpage}

\newpage
\begin{figure*}
\hspace{-2.5cm}
\includegraphics[scale = 0.65, trim = 25 0 65 0, clip]{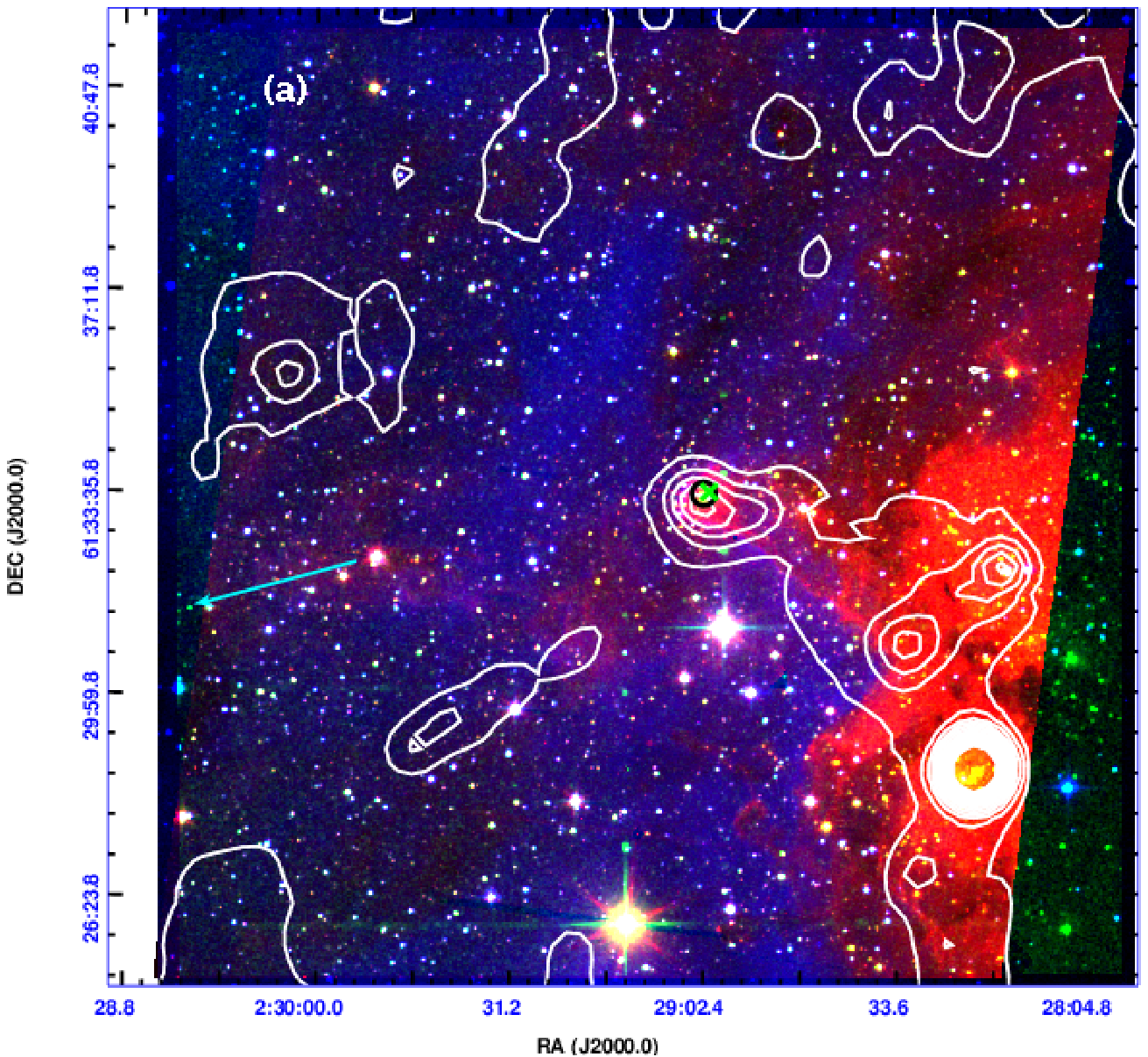}
\includegraphics[scale =0.90, trim = 50 0 90  0, clip]{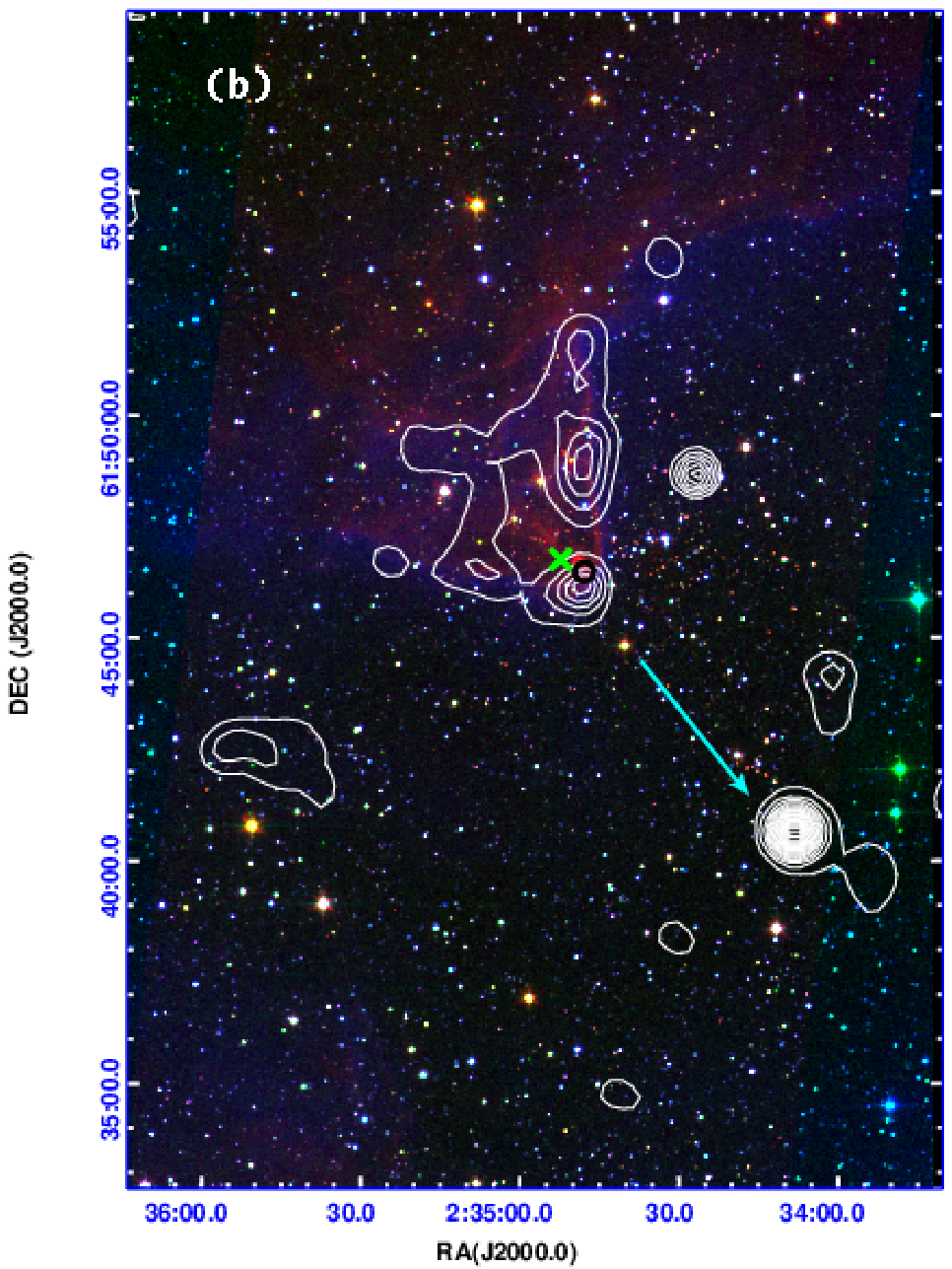}
\includegraphics[scale = 0.6, trim = 0 0 0  0, clip]{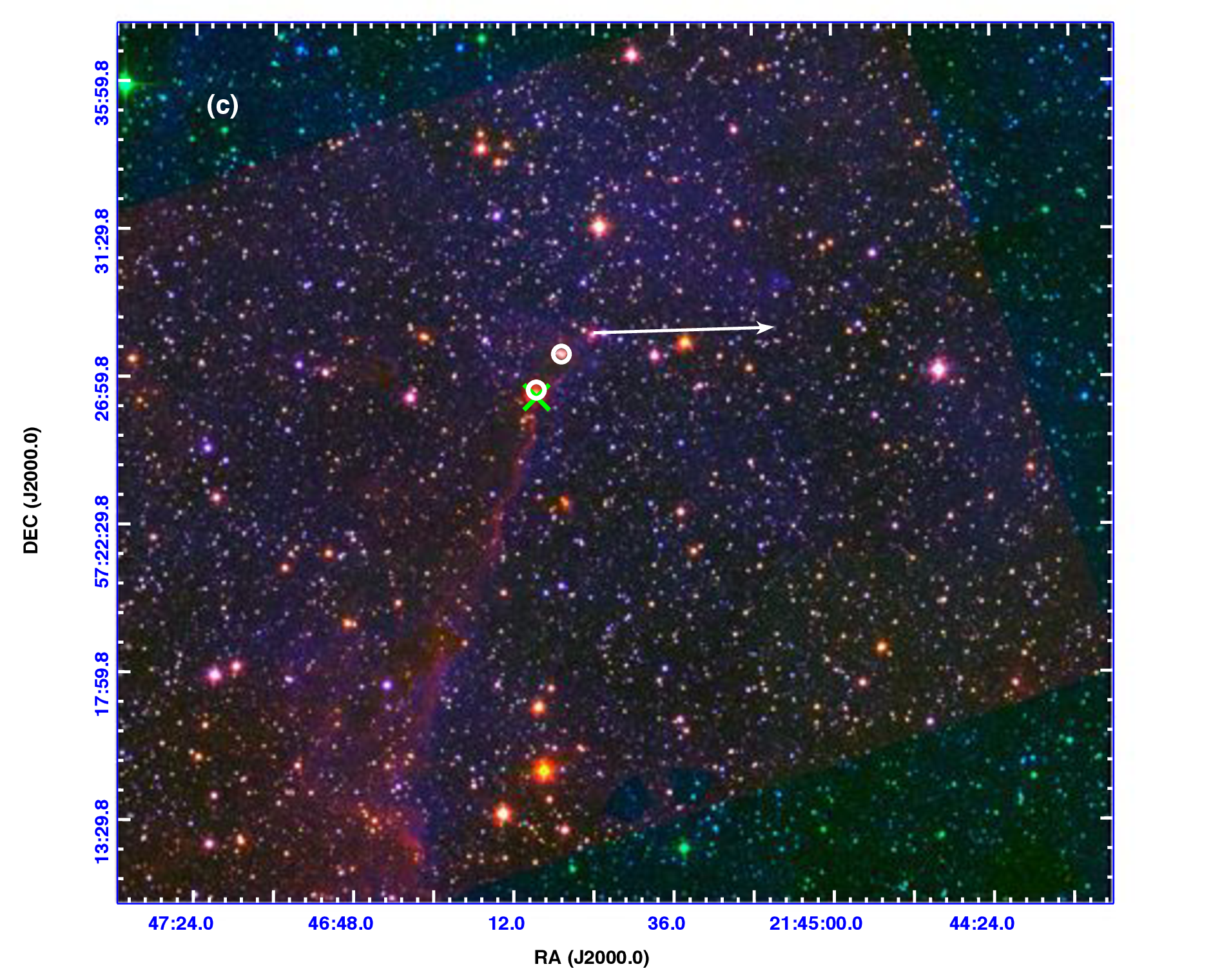}
\caption{DSS2-R (blue), 2MASS-Ks (green) and IRAC-4.5 $\micron$ (red) colour-composite image of (a)~BRC\,5, (b)~BRC\,7, and (c)~BRC\,39. 
Open circles mark the submillimeter cores, and crosses mark the IRAS sources. In BRC\,5 and BRC\,7, contours represent the NVSS 1.4~GHz continuum 
emission. The arrow shows the direction of the ionising star(s).}
\label{fig3}
\end{figure*}

\newpage

\begin{figure*}
\centering
\includegraphics[scale = .3, trim = 5 5 5  5, clip]{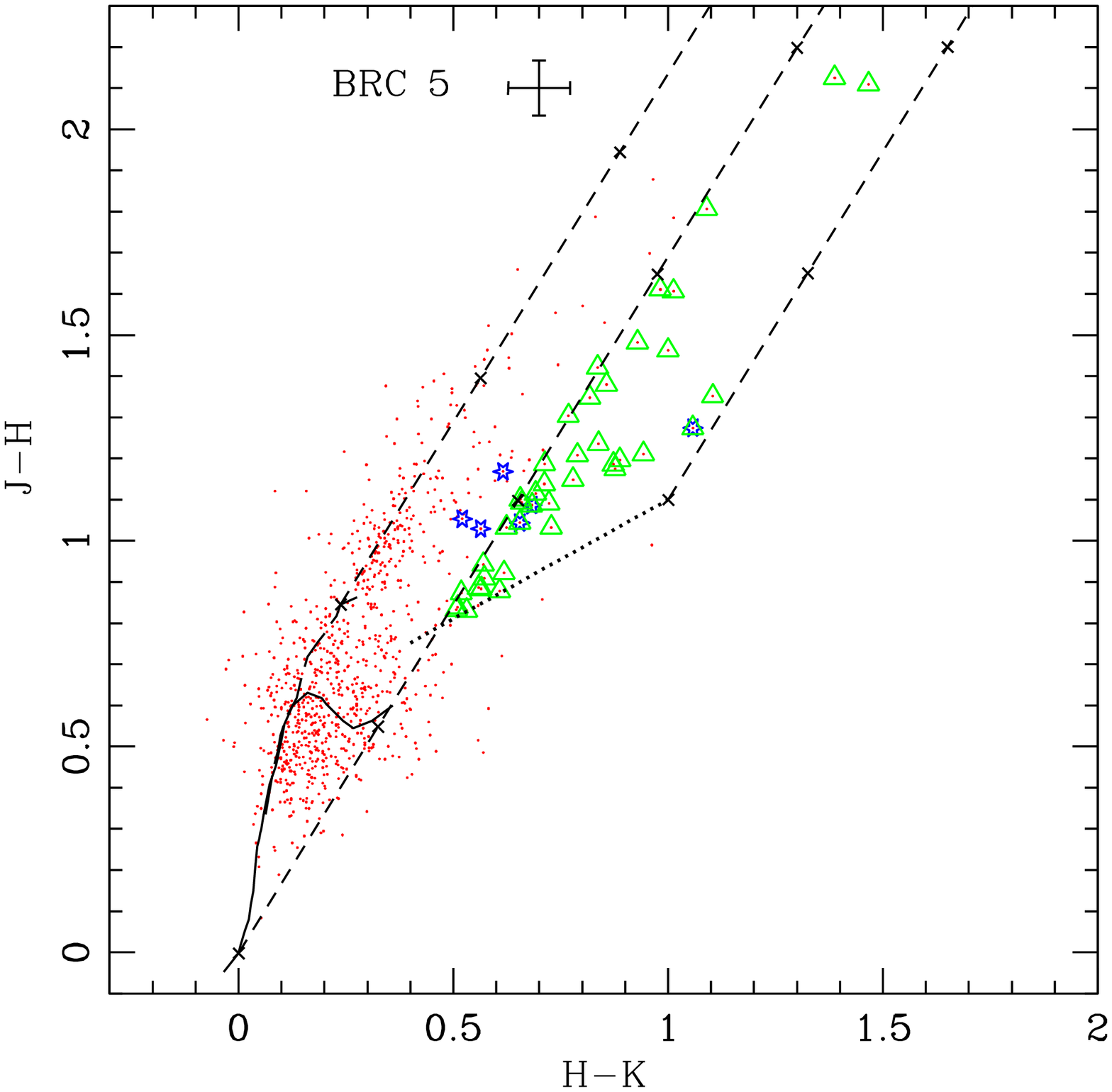}
\includegraphics[scale = .3, trim = 5 5 5  5, clip]{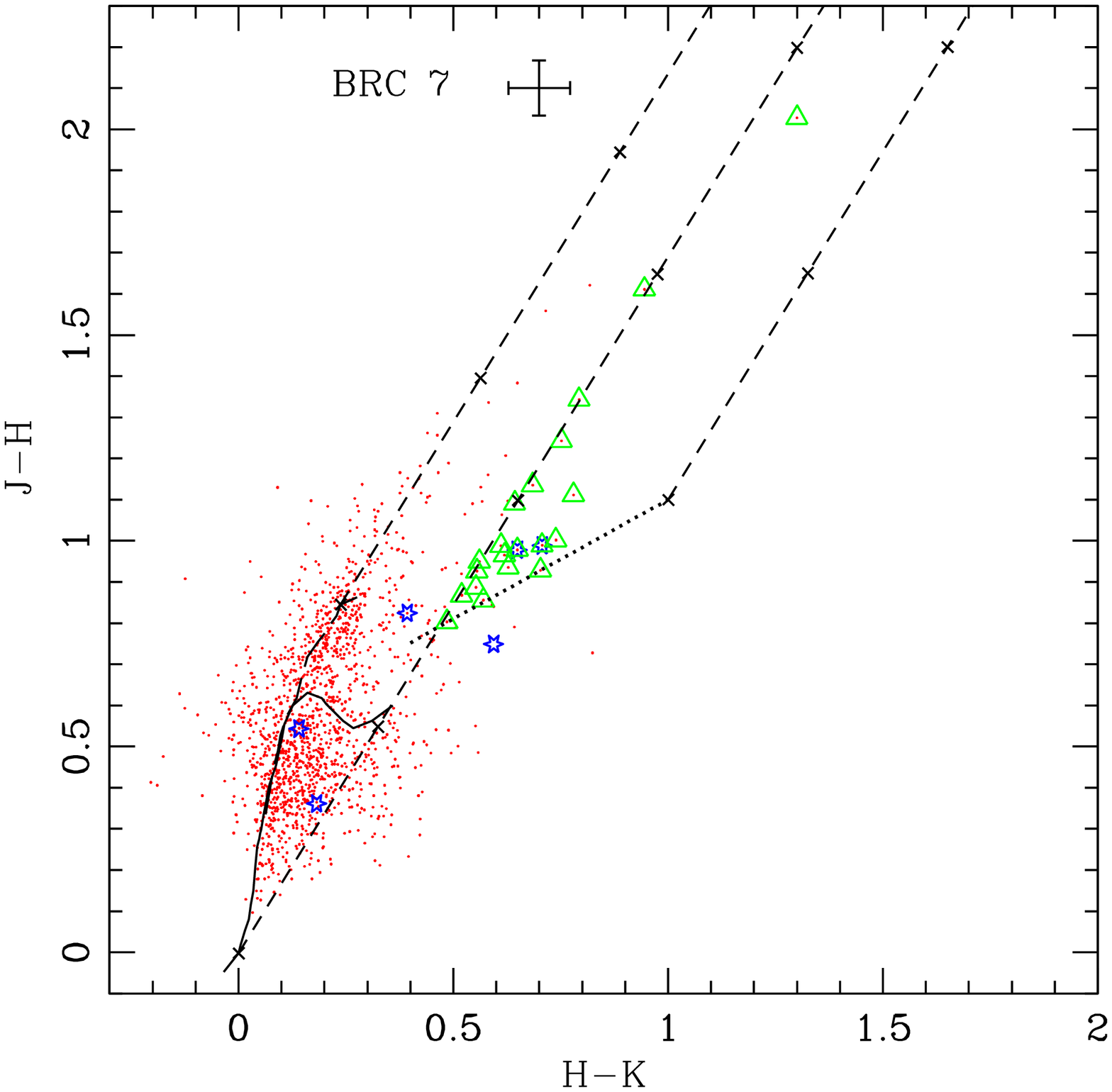}
\includegraphics[scale = .3, trim = 5 5 5  5, clip]{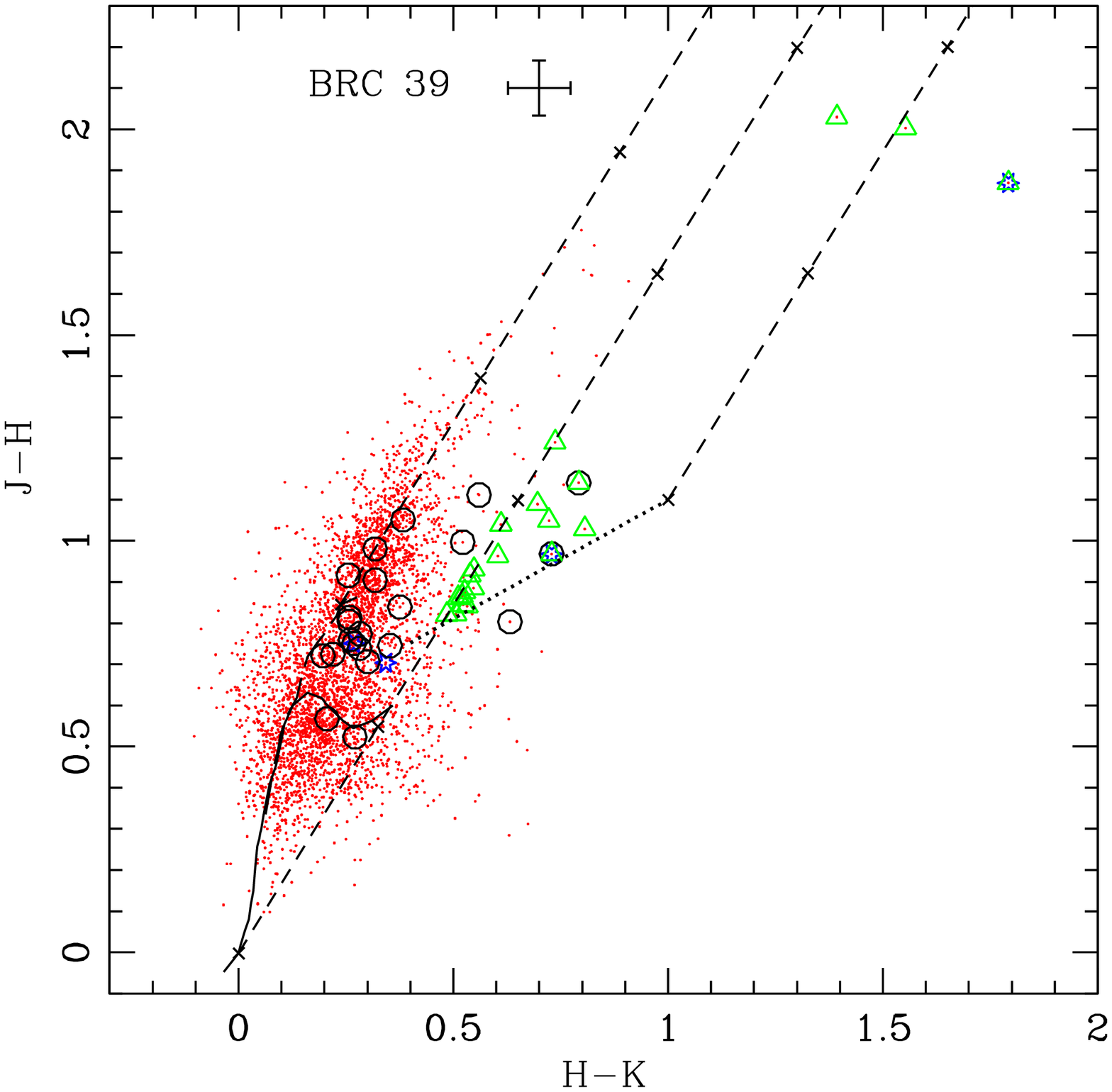}
\caption{$ (J - H)/(H - K)$ NIR colour-colour diagrams for BRC\,5, BRC\,7 and BRC\,39. 
Open triangles represent NIR excess stars and blue asterisk symbols represent the H$\alpha$ emission stars (Ogura et al. 2002). 
In BRC\,39, the large open circles represent H$\alpha$ emission stars (Nakano et al. 2012).  The error bars in the top 
middle show average errors in the colours. The continuous and dashed curves represent the unreddened main-sequence and giant 
loci (Bessell $\&$ Brett 1988), respectively. The dotted line represents the intrinsic colours of CTTSs (Meyer et al. 1997) and  
the parallel dashed lines are the reddening vectors with crosses separated by $A_V$ $=$ 5 mag. }
\label{fig1}
\end{figure*}
\newpage
\begin{figure*}
\centering
\includegraphics[scale = .3, trim = 5 5 5  5, clip]{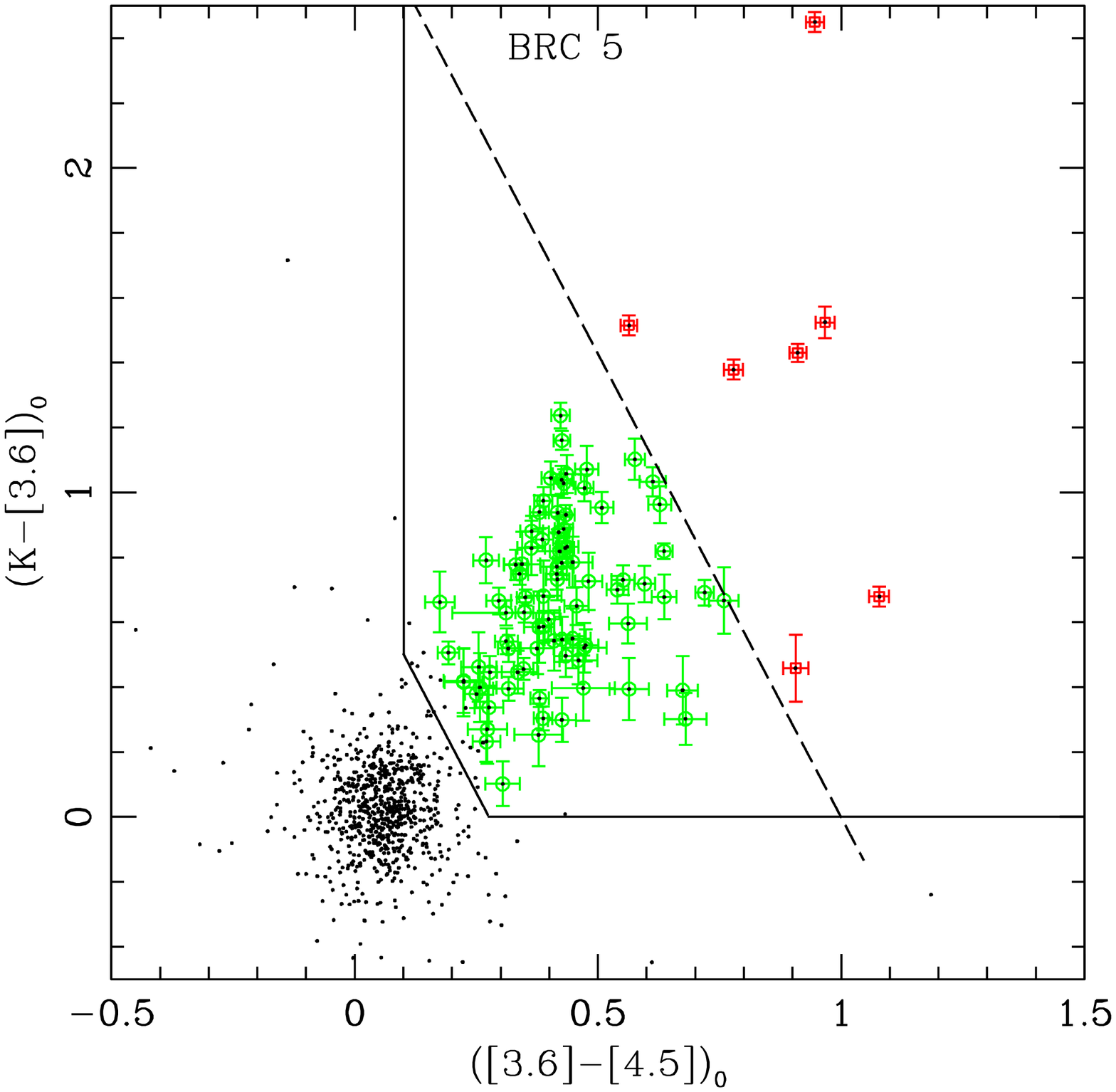}
\includegraphics[scale = .3, trim = 5 5 5  5, clip]{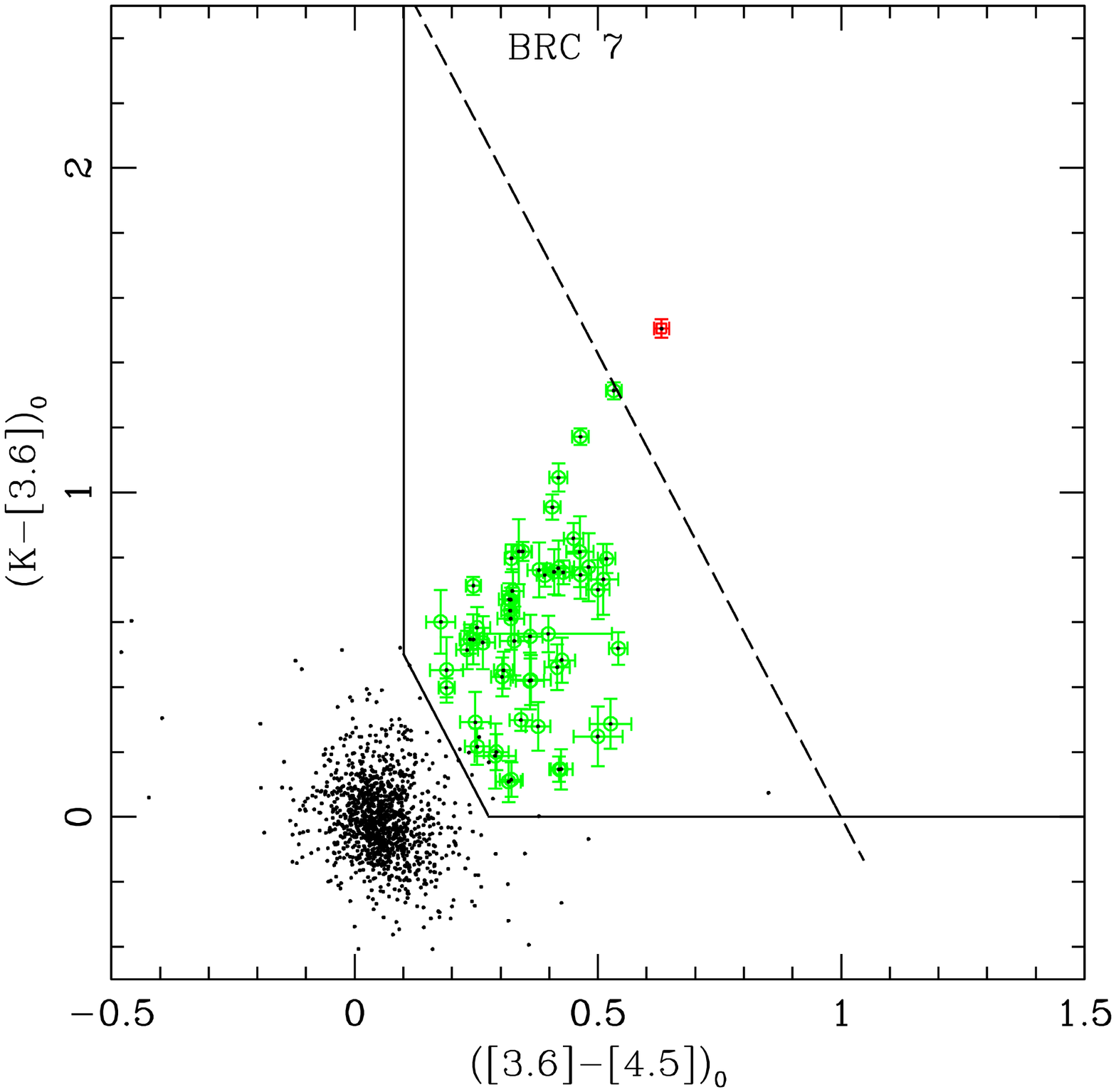}
\includegraphics[scale = .3, trim = 5 5 5  5, clip]{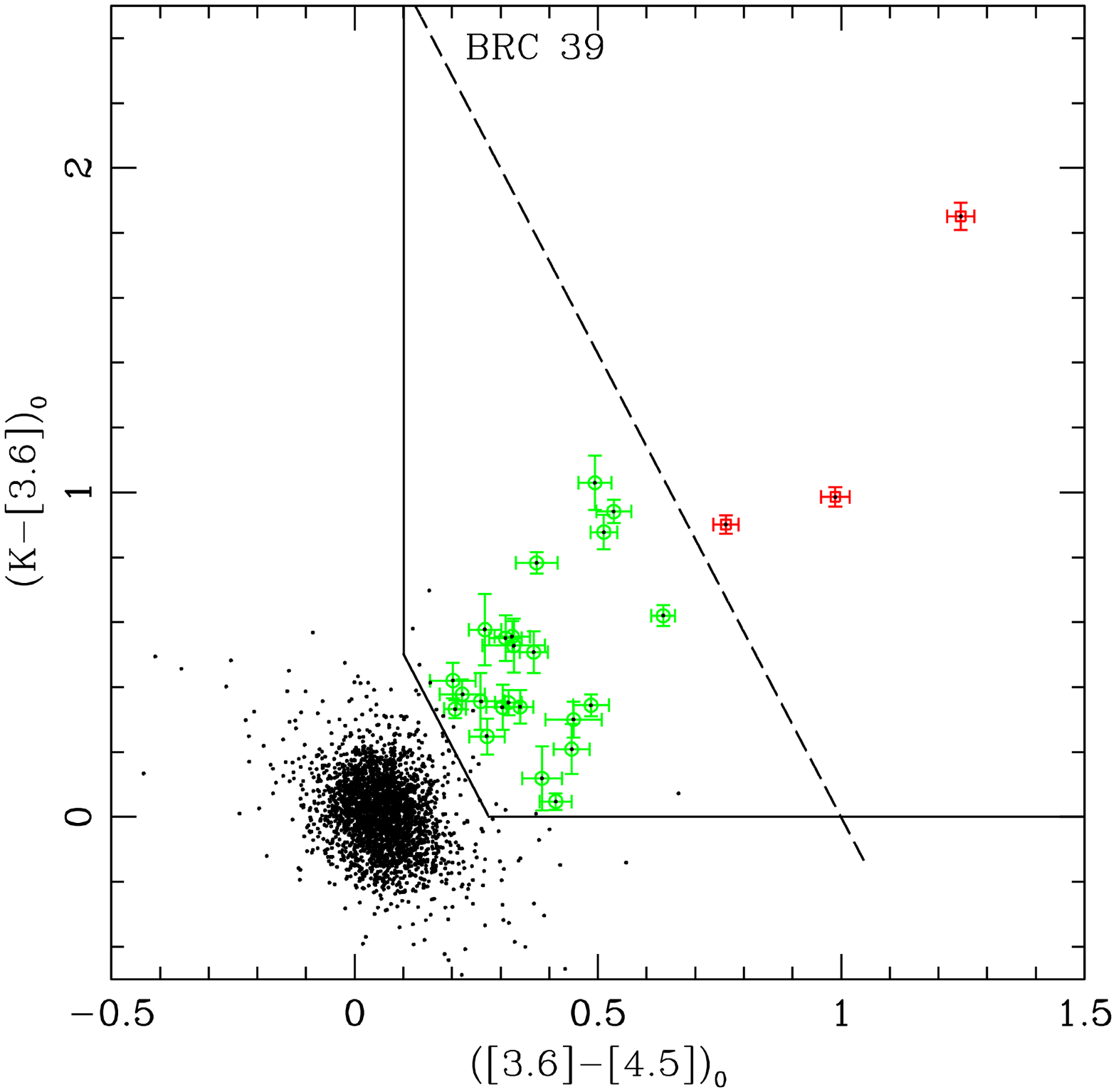}

\caption{$2MASS-IRAC$ colour-colour diagrams for BRC\,5, BRC\,7 and BRC\,39. 
Red squares are Class I sources and green circles are Class II sources. The dashed lines and continuous lines are 
the criteria used by Gutermuth et al. (2009) to separate Class II sources from field stars/MS sources and from Class I sources, 
respectively. The error bars show the errors in the respective colours of Class I and Class II sources. } \label{fig2}
\end{figure*}


\newpage
\begin{figure*}
\centering
\includegraphics[scale = .35, trim = 5 5 5  5, clip]{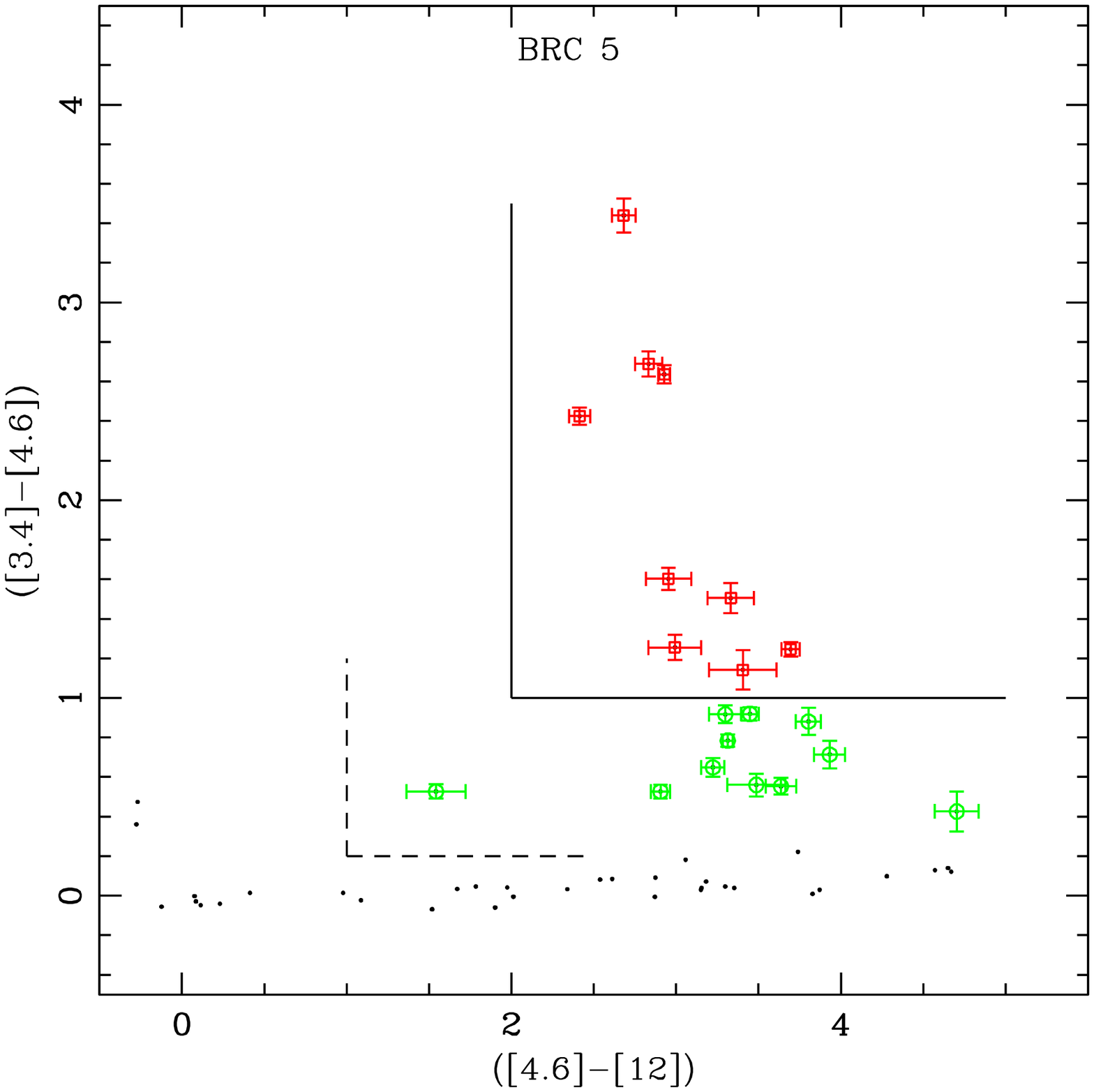}
\includegraphics[scale = .35, trim = 5 5 5  5, clip]{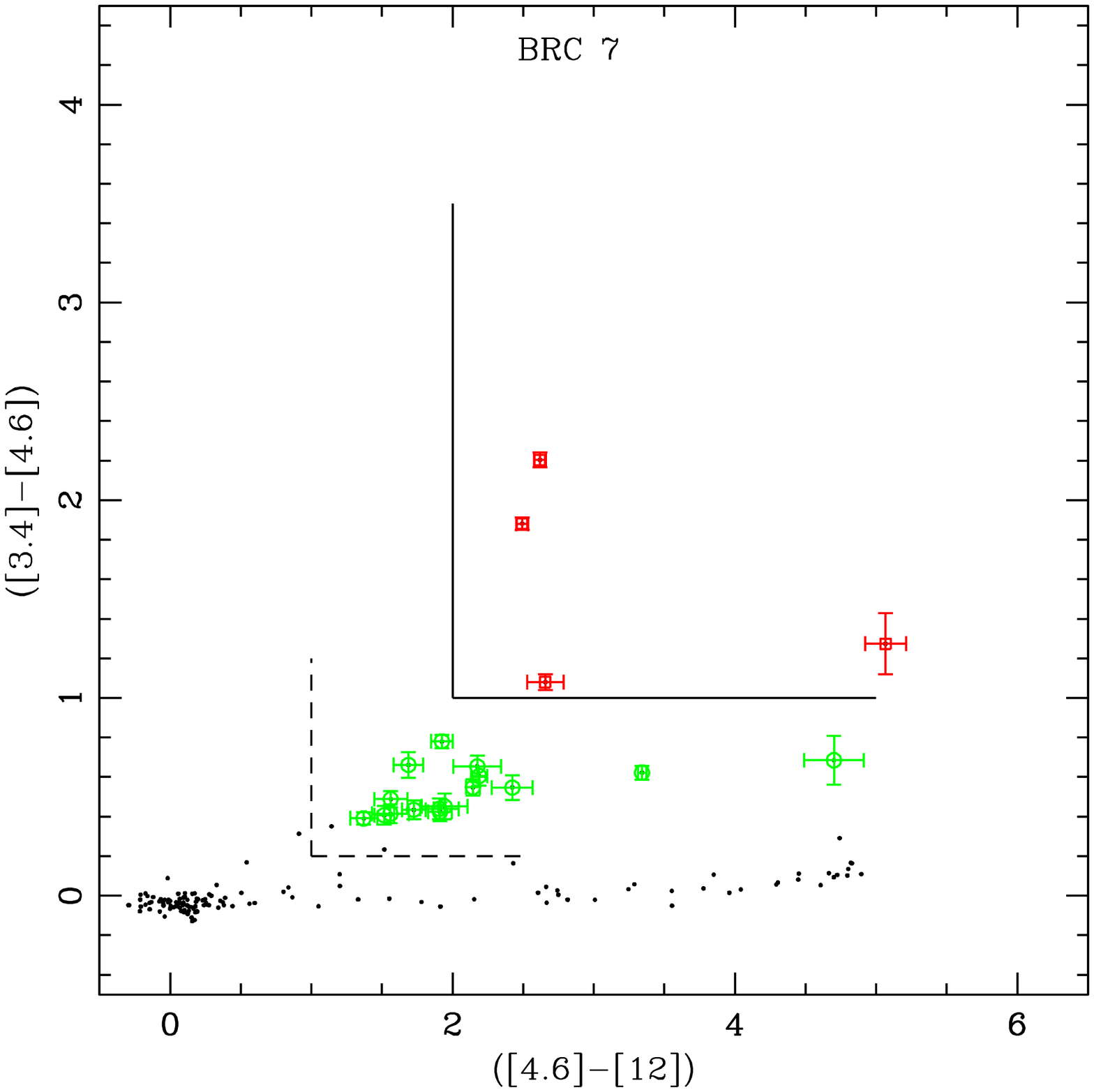}
\includegraphics[scale = .35, trim = 5 5 5  5, clip]{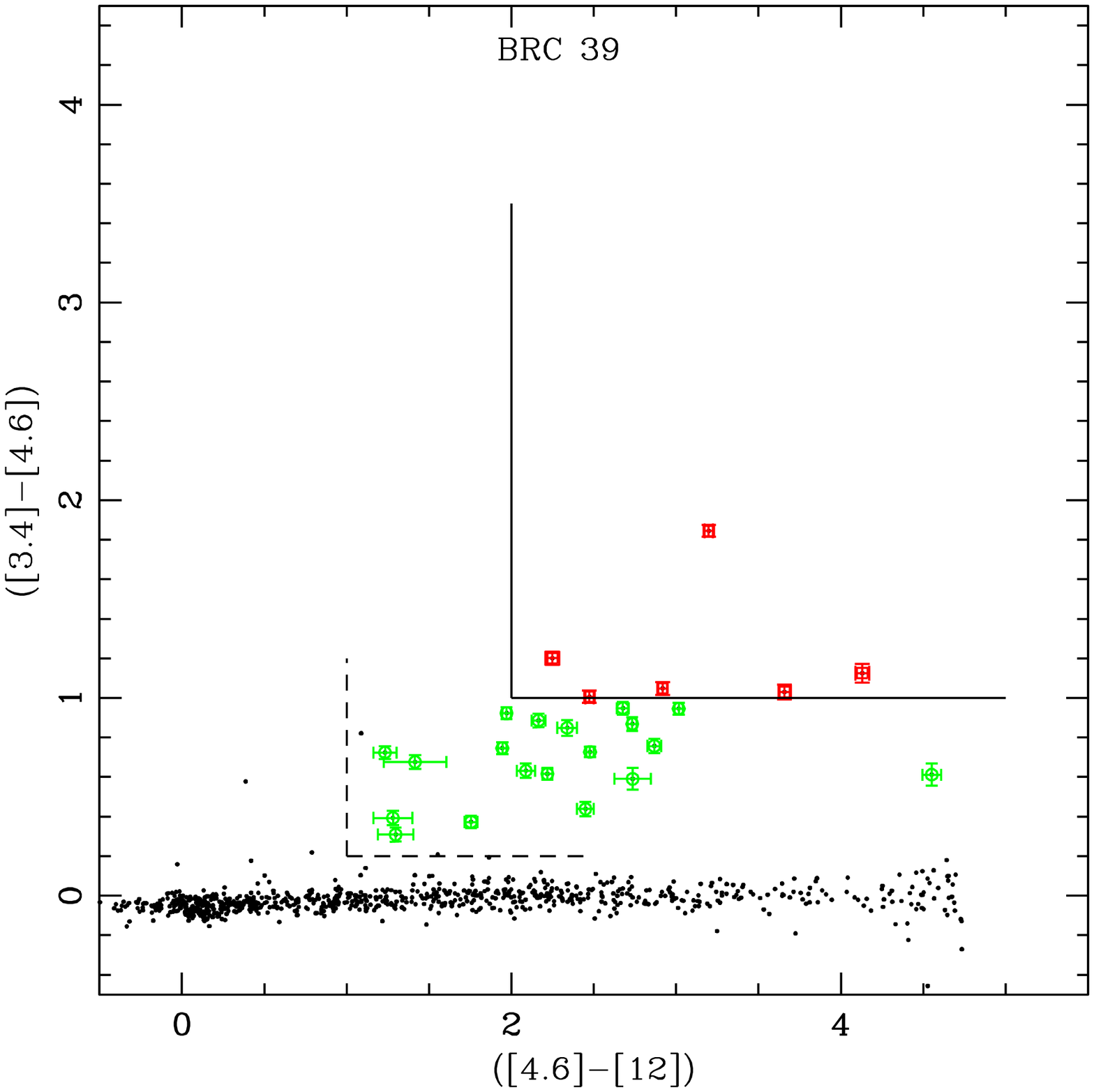}
\caption{The WISE colour-colour diagrams for the sources in BRC\,5, BRC\,7 and BRC\,39. Green circles represent 
Class II sources and red squares represent Class I sources.  The dashed lines and continuous lines are the criteria used by 
Koenig et al. (2012) to separate Class II sources from field/MS sources and from Class I sources, respectively.  
The error bars show the errors in the respective colours of the identified Class I and Class II sources. }
\label{fig4}
\end{figure*}

\begin{figure*}
\centering
\includegraphics[scale = .5, trim = 5 5 5  5, clip]{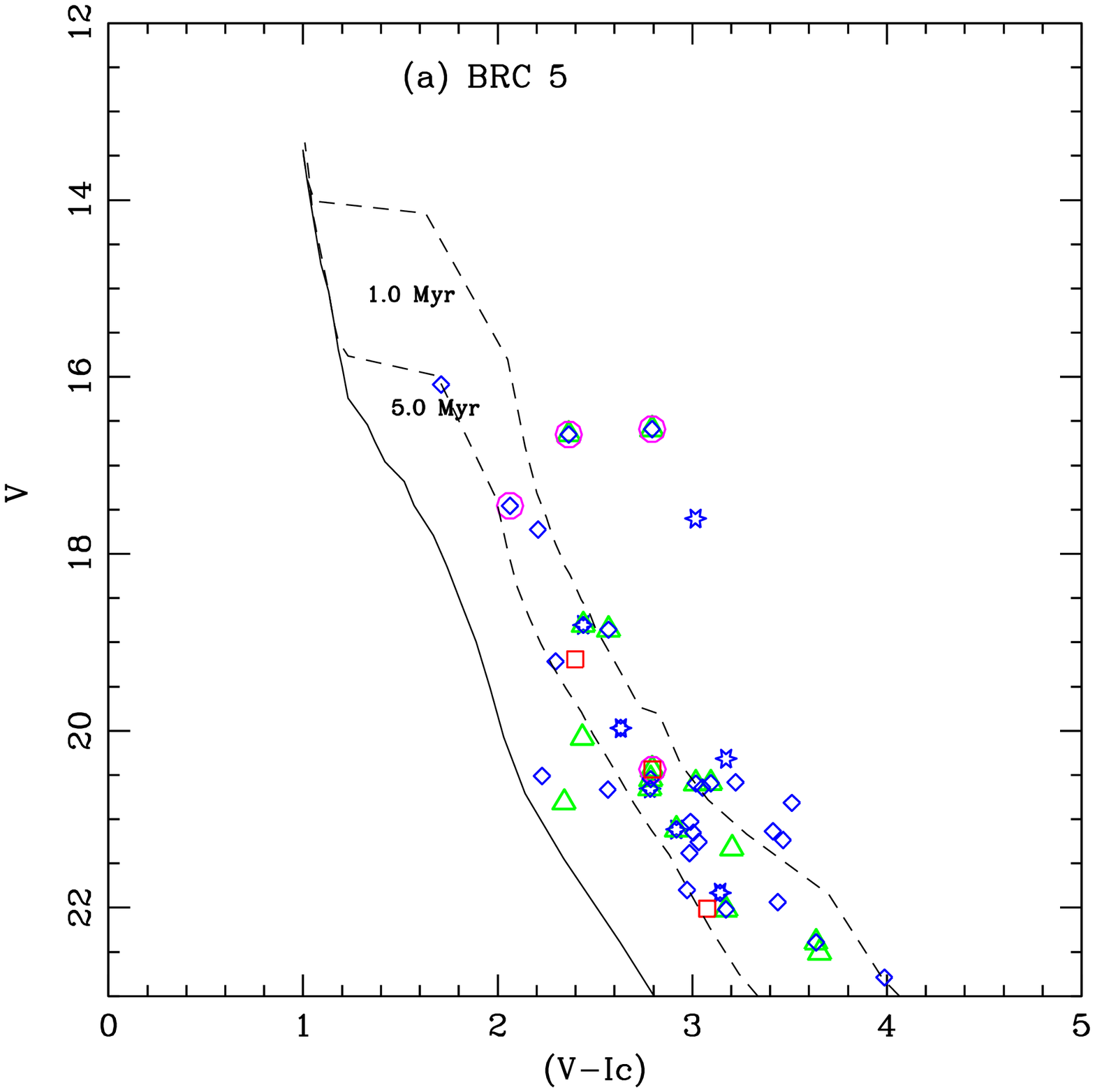}
\includegraphics[scale = .5, trim = 5 5 5  5, clip]{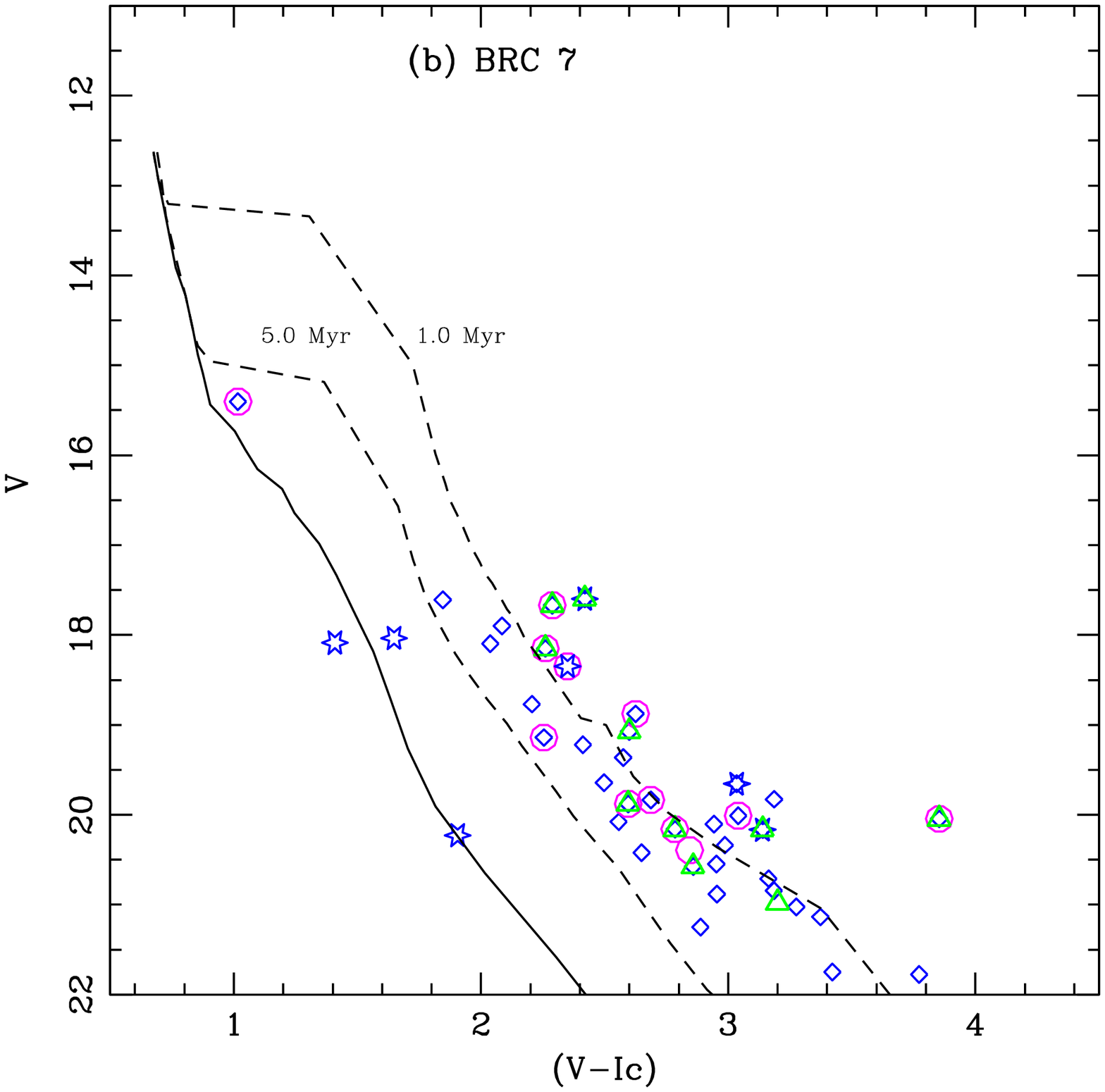}
\caption{Optical $V/(V - I_c)$ colour-magnitude diagrams of (a)~BRC\,5 and (b)~BRC\,7. Open triangles represent NIR excess sources 
identified by 2MASS data, squares and diamonds represent the IRAC Class~I and Class~II sources respectively. 
Class~II sources selected based on the WISE colours are shown as open circles. 
The asterisk symbols represent the positions of H$\alpha$ emission stars (Ogura et al. 2002).  The continuous line 
is the zero-age MS and dashed lines are the 1-Myr and 5-Myr PMS isochrones (Siess et al. 2000).}
\label{fig6}
\end{figure*}

\newpage

\begin{figure*}
\centering
\includegraphics[scale = .5, trim = 5 5 5  5, clip]{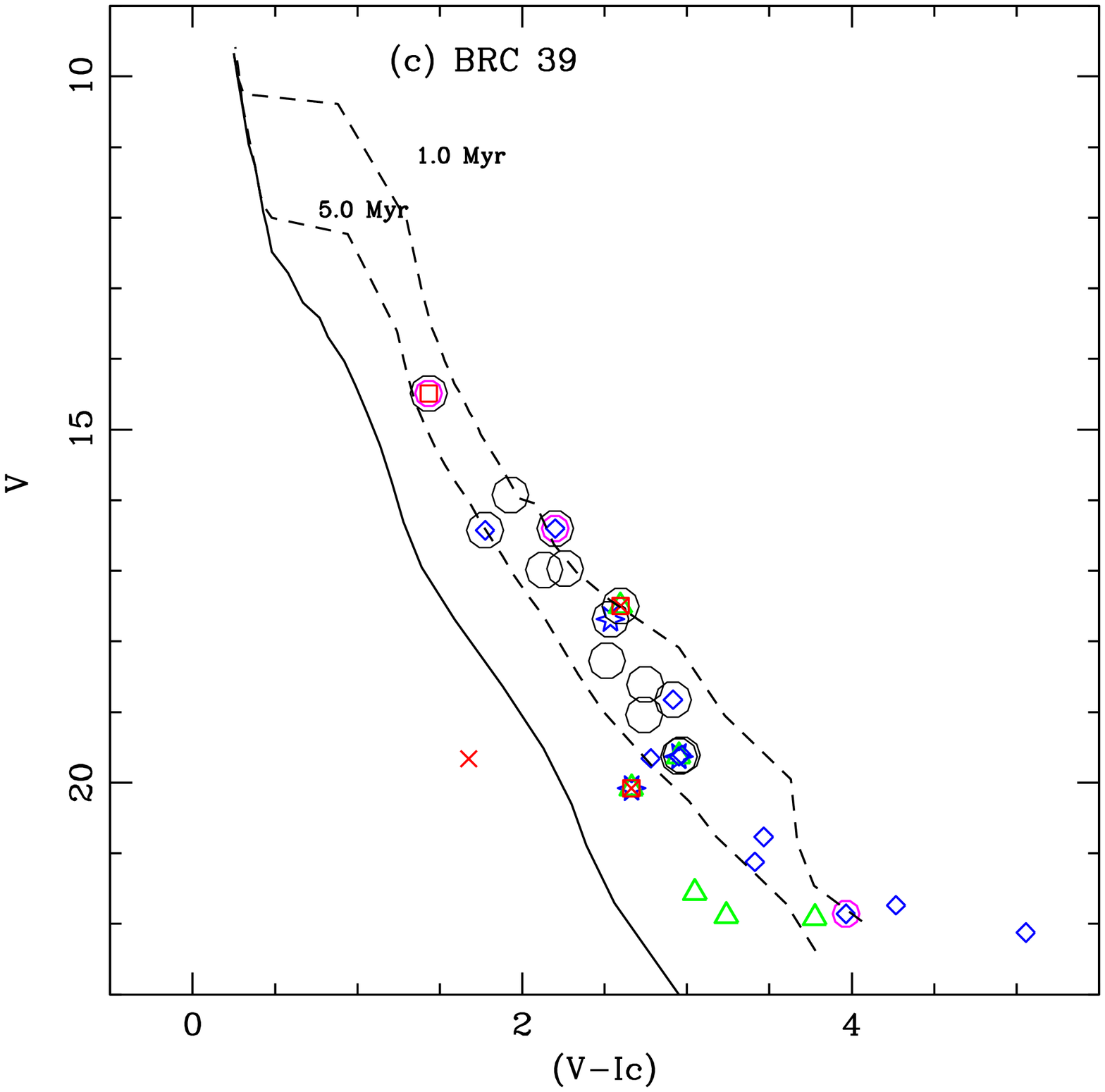}
{\hspace{16cm}Fig. \ref{fig6} Cont. colour-magnitude diagram of the identified YSOs in BRC 39 region. Large open circles are the H$\alpha$ sources from Nakano et al. (2012) and crosses are Class~I sources selected based on the WISE colours. Other symbols are same as in previous figures. }
\end{figure*}
\begin{figure*}
\centering
\includegraphics[scale = .55, trim = 0 0 0  0, clip]{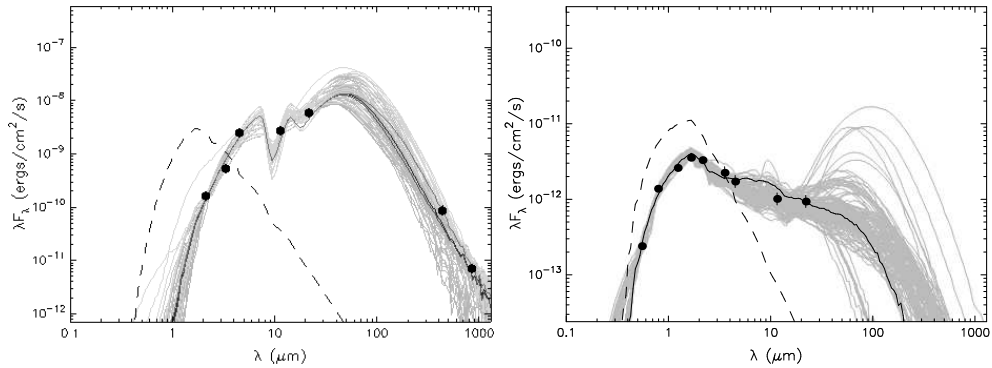}
\includegraphics[scale = .55, trim = 0 0 0  0, clip]{sed_BRC7_17.eps}
\caption{Sample SEDs for Class I (left panel) and Class II (right panel) sources in BRC\,5, and BRC\,7, respectively.  
The dashed line shows the stellar photosphere corresponding to the central source of the best-fitting model.  
The grey lines show subsequent good fits to the data points, and the black line shows the best fit.  
The filled circles denote the input flux values. }
\end{figure*} 

\newpage
\begin{figure*}
\hspace{-2.4cm}
\includegraphics[scale = 0.65, trim = 15 0  20  0, clip]{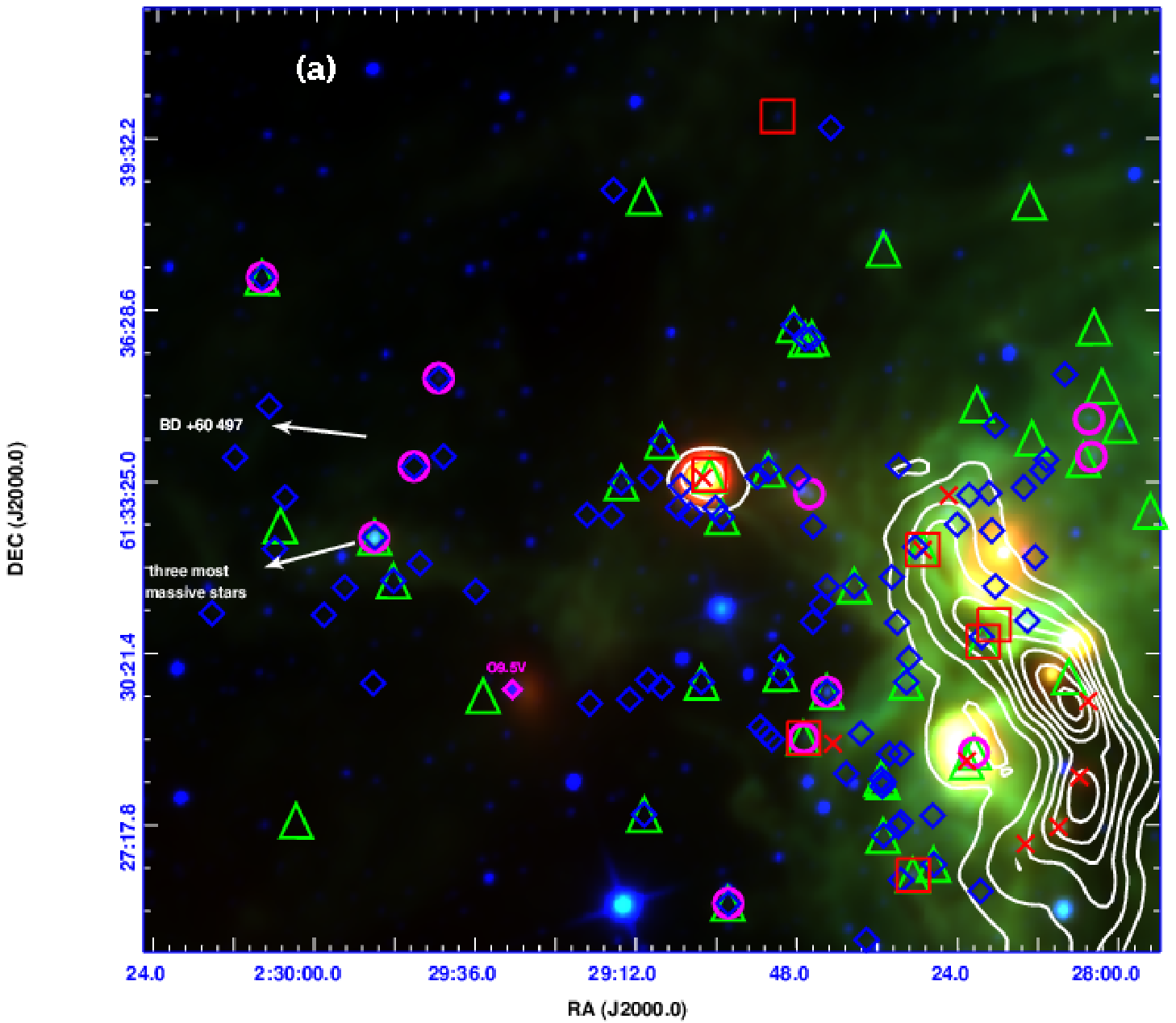}
\label{fig5a}
\includegraphics[scale = 0.65, trim = 20 5 50  0, clip]{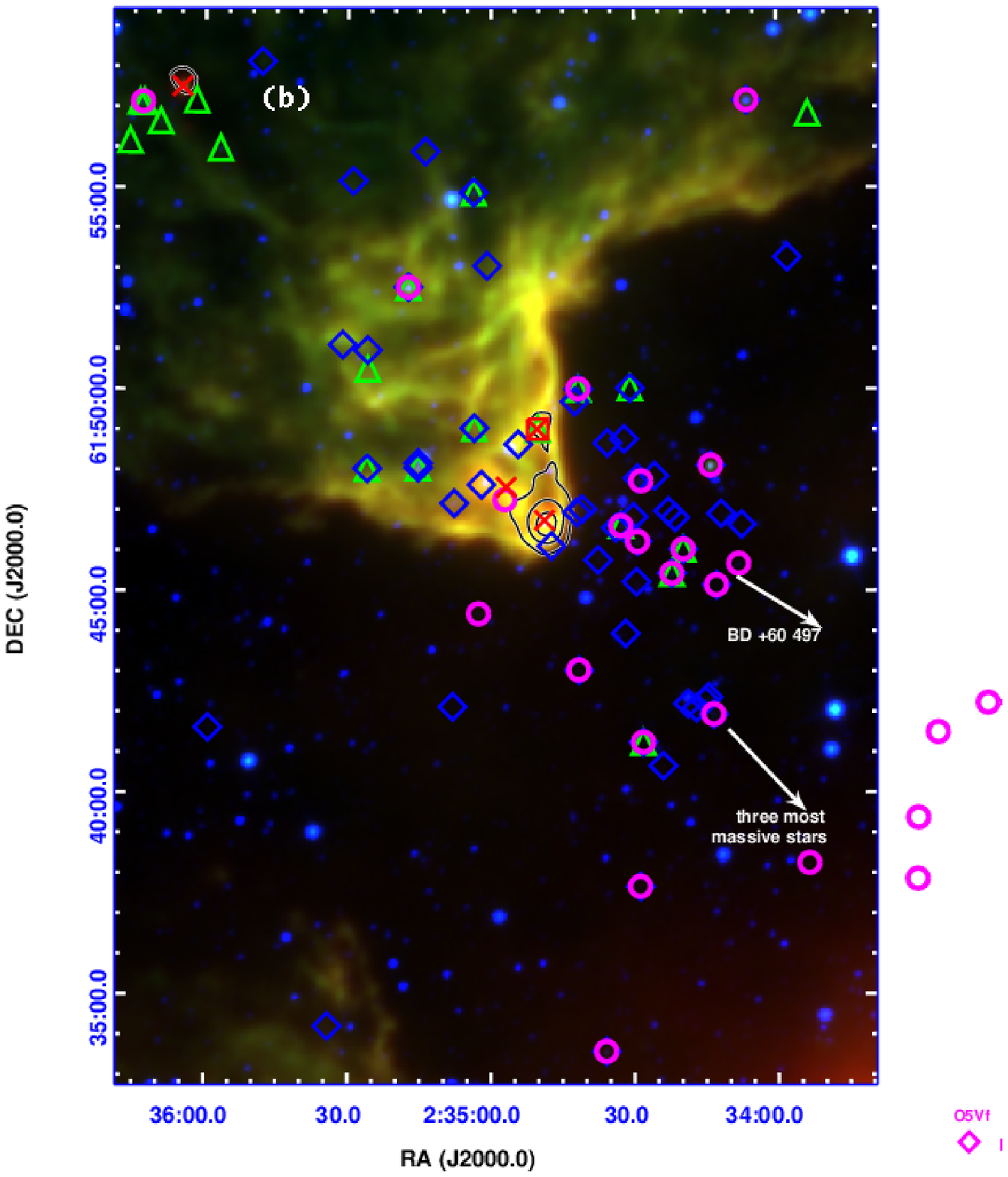}
\includegraphics[scale = 0.70, trim = 0 0  25  5, clip]{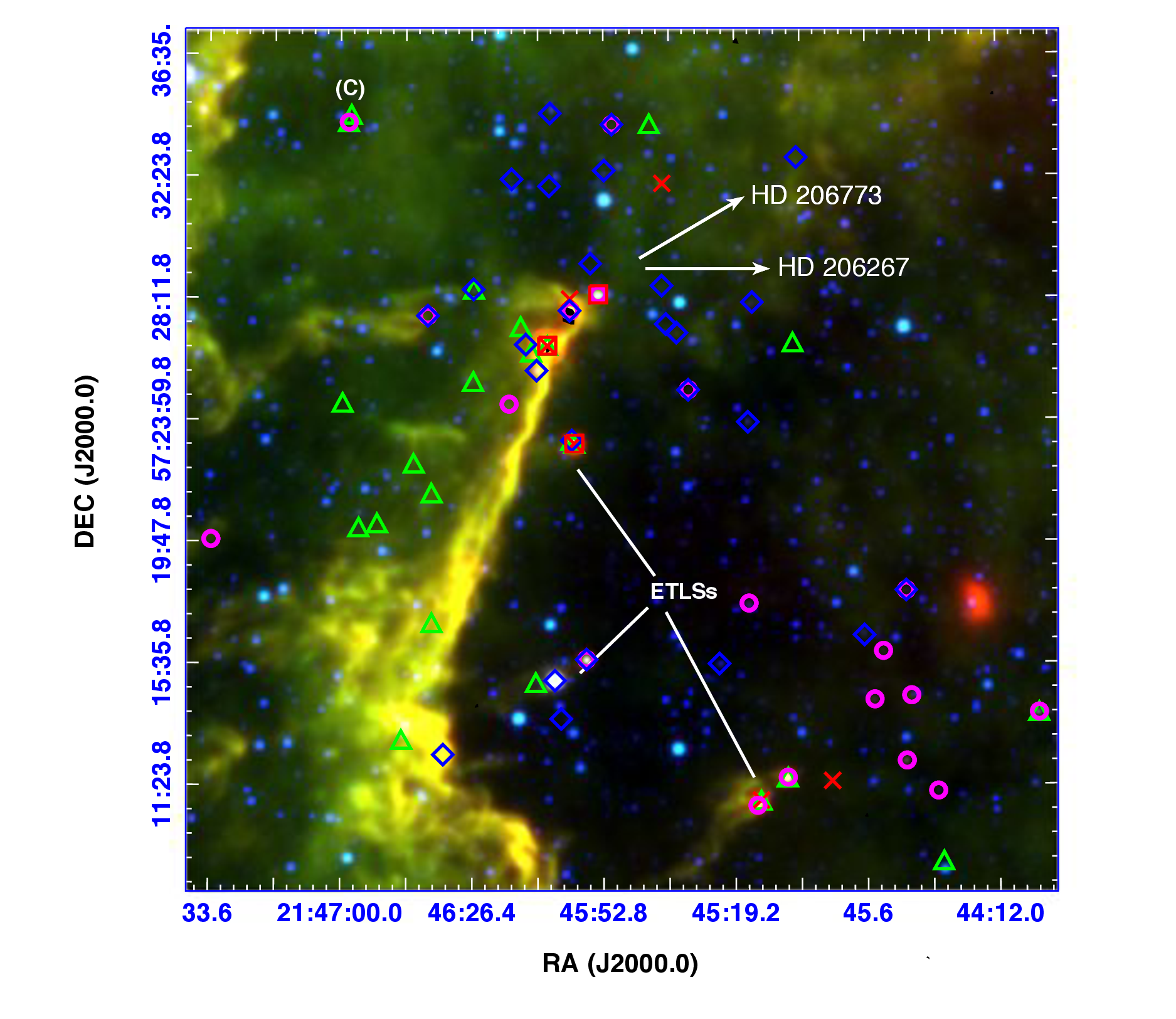}
\caption{ Spatial distribution of the identified YSOs in (a)~BRC\,5, (b)~BRC\,7, (c)~BRC\,39 overplotted on the 
WISE three-colour composite image at 22~$\micron$ (in red), 12~$\micron$ (in green), and 3.4~$\micron$ (in blue).  
Open triangles represent NIR excess sources from 2MASS, squares and diamonds represent the IRAC Class~I 
and Class~II sources respectively. Class~I and Class~II sources selected based on the WISE colours are
 shown as crosses and open circles, respectively.  The contours represent the 1.1~mm emission obtained 
from the CSO BOLOCAM images.  The elephant trunk-like structures seen toward BRC\,39 are marked. }
\label{fig5b}
\end{figure*}


\end{document}